\documentclass[11pt,a4paper]{article}

\usepackage{jheppub}
\usepackage{graphicx}
\usepackage{amsmath}
\usepackage{placeins}
\usepackage{multirow}
\usepackage{braket}
\usepackage{xcolor}
\usepackage{wrapfig}

\graphicspath{{figs/}}

\renewcommand\thesection{\arabic{section}}

\newcommand{\beq}{\begin{eqnarray}}
\newcommand{\eeq}{\end{eqnarray}}
\newcommand{\beqnn}{\begin{eqnarray*}}
\newcommand{\eeqnn}{\end{eqnarray*}}
\newcommand{\Tr}{\ensuremath{\mathrm{Tr}}}
\newcommand{\Var}{\ensuremath{\mathrm{Var}}}
\newcommand{\Eff}{C^{\rm (eff)}}
\newcommand{\CP}{\mathrm{CP}}
\newcommand{\SU}{\mathrm{SU}}
\newcommand{\cool}{\mathrm{cool}}
\newcommand{\dd}{{\mathrm{d}}}
\newcommand{\nstep}{n_{\mathrm{step}}}
\newcommand{\nev}{n_{\mathrm{ev}}}

\newcommand{\ndof}{n_{\mathrm{dof}}}
\newcommand{\nbetween}{n_{\mathrm{between}}}
\newcommand{\tauint}{\tau_{\mathrm{int}}}

\newcommand{\DKL}{\tilde{D}_{\mathrm{KL}}}
\newcommand{\ESS}{\mathrm{ESS}}
\newcommand{\hESS}{\hat{\mathrm{ESS}}}

\newcommand{\clov}{\mathrm{clov}}
\newcommand{\chiL}{\chi_{\scriptscriptstyle{\rm L}}}
\newcommand{\QL}{Q_{\scriptscriptstyle{\rm L}}}

\newcommand{\neavg}[1]{\langle #1 \rangle_\mathrm{f}}
\newcommand{\neavgn}[1]{\langle #1 \rangle_{\mathrm{f}, \lambda(n)}}
\newcommand{\eqavg}[1]{\langle #1 \rangle_{\mathrm{eq}, \lambda(n)}}
\newcommand{\pd}[2]{\frac{\partial {#1}}{\partial {#2}}}

\newcommand{\Pf}{\mathcal{P}_{\mathrm{f}}}
\newcommand{\Pre}{\mathcal{P}_{\mathrm{r}}}
\newcommand{\Wd}{W_{\mathrm{d}}}
\newcommand{\U}{\mathcal{U}}

\begin{document}

\title{Scaling flow-based approaches for topology sampling in $\SU(3)$ gauge theory}

\author[a]{Claudio Bonanno,}
\affiliation[a]{Instituto de F\'isica Te\'orica UAM-CSIC, c/ Nicol\'as Cabrera 13-15, Universidad Aut\'onoma de Madrid, Cantoblanco, E-28049 Madrid, Spain}
\emailAdd{claudio.bonanno@csic.es}

\author[b,c]{Andrea Bulgarelli,}
\affiliation[b]{Dipartimento di Fisica,  Universit\'a degli Studi di Torino and INFN, Sezione di Torino, Via Pietro Giuria 1, I-10125 Turin, Italy}
\emailAdd{andrea.bulgarelli@unito.it}
\affiliation[c]{Transdisciplinary Research Area ``Building Blocks of Matter and Fundamental Interactions'' (TRA Matter) and Helmholtz Institute for Radiation and Nuclear Physics (HISKP), University of Bonn, Nussallee 14-16, 53115 Bonn, Germany}

\author[b,d]{Elia Cellini,}
\emailAdd{elia.cellini@ed.ac.uk}
\affiliation[d]{Higgs Centre for Theoretical Physics, School of Physics and Astronomy,\\
The University of Edinburgh, Edinburgh EH9 3FD, United Kingdom}
\author[b]{Alessandro Nada,}
\emailAdd{alessandro.nada@unito.it}

\author[b]{Dario Panfalone,}
\emailAdd{dario.panfalone@unito.it}

\author[e]{Davide Vadacchino,}
\affiliation[e]{Centre for Mathematical Sciences, University of Plymouth, Plymouth, PL4 8AA, United Kingdom}
\emailAdd{davide.vadacchino@plymouth.ac.uk}

\author[b]{Lorenzo Verzichelli}
\emailAdd{lorenzo.verzichelli@unito.it}

\abstract{We develop a methodology based on out-of-equilibrium simulations to mitigate topological freezing when approaching the continuum limit of lattice gauge theories. We reduce the autocorrelation of the topological charge employing open boundary conditions, while removing exactly their unphysical effects using a non-equilibrium Monte Carlo approach in which periodic boundary conditions are gradually switched on. We perform a detailed analysis of the computational costs of this strategy in the case of the four-dimensional $\mathrm{SU}(3)$ Yang-Mills theory. After achieving full control of the scaling, we outline a clear strategy to sample topology efficiently in the continuum limit, which we check at lattice spacings as small as $0.045$ fm. We also generalize this approach by designing a customized Stochastic Normalizing Flow for evolutions in the boundary conditions, obtaining superior performances with respect to the purely stochastic non-equilibrium approach, and paving the way for more efficient future flow-based solutions.}


\maketitle
\flushbottom

\section{Introduction}\label{sec:intro}

Numerical Markov Chain Monte Carlo (MCMC) simulations of lattice field theories are amongst the most powerful tools for exploring the non-perturbative regime of non-Abelian gauge theories. Over the past decades, their use has provided first-principles insights into the theoretical and phenomenological properties of several lattice-regularized models, the most prominent example being lattice Quantum Chromodynamics (QCD). Nonetheless, this approach is accompanied by a number of highly non-trivial computational challenges. Although advances in the architecture of supercomputing machines have greatly expanded the range of feasible calculations, the development of more efficient and sophisticated algorithms remains essential to overcome these limitations.

In lattice gauge theories, and in particular in lattice QCD, one of the most severe numerical issues within the MCMC framework is the so-called \emph{critical slowing down}, in particular that of topological modes. As the continuum limit is approached, the computational cost required to obtain statistically independent configurations grows rapidly with decreasing lattice spacing, ultimately leading to a loss of ergodicity of the Markov chain. This is a critical issue, since ergodicity is a key assumption underlying the validity of ensemble averages as estimators of expectation values. 
For most observables, critical slowing down manifests as a polynomial growth of the autocorrelation time with the inverse lattice spacing with a small exponent. In contrast, for topological quantities such as the topological charge $Q$~\cite{Alles:1996vn,DelDebbio:2002xa,DelDebbio:2004xh,Schaefer:2010hu}, the scaling is found to be much more severe and compatible with a polynomial with a large exponent or even with an exponential. This can be understood in terms of the MCMC dynamics of topological modes when standard local updating algorithms are adopted to generate the Markov chain: while for non-topological quantities this is essentially diffusive, for topological ones this is dominated by jumps over the potential barriers among different pseudo-topological sectors. Such barriers eventually diverge in the continuum limit to restore a proper notion of topological winding number~\cite{Luscher:1981zq}. Since no change of topological sector is allowed via a local deformation of the gauge fields, it becomes increasingly difficult to change the winding number of a given lattice gauge field as the lattice spacing approaches zero. This severe ergodicity problem affecting the sampling of the topological charge typically results, on fine lattices, in few or even no fluctuations of $Q$ during feasible MCMC histories: for this reason it is typically called \emph{topological freezing}.

Topological freezing poses a serious problem for the determination of topological quantities from lattice simulations, most notably the topological susceptibility, a quantity of the utmost theoretical and phenomenological importance which has been widely addressed in the lattice literature~\cite{Durr:2006ky, Ce:2015qha,
Bonati:2015sqt, Athenodorou:2020ani, Bonanno:2023ple, Durr:2025qtq,DelDebbio:2004ns, Luscher:2010ik, Cichy:2015jra, Bonanno:2019xhg,Ce:2016awn, Bonati:2016tvi, Bonanno:2020hht, Athenodorou:2021qvs}. 
However, such a severe loss of ergodicity can in principle bias any expectation value estimated from topologically-frozen samples. It is well-known, for instance, that it can affect the calculation of particle spectra~\cite{Brower:2003yx, Aoki:2007ka}, as well as observables computed after the gradient flow like the action density~\cite{Fritzsch:2013yxa,RamosMartinez:2023tvx}, necessary to obtain the reference scale $t_0$ or the renormalized strong coupling. For this reason, mitigating topological freezing is not only crucial for studies of topological quantities, but also to ensure the reliability of a wide range of lattice results. Developing new numerical strategies to address this issue is a major focus within the lattice community, leading to substantial progress in the last decade~\cite{Bietenholz:2015rsa,Laio:2015era,Luscher:2017cjh,Bonati:2017woi,Giusti:2018cmp,Florio:2019nte,Funcke:2019zna,Albandea:2021lvl,Cossu:2021bgn,Borsanyi:2021gqg,Fritzsch:2021klm,Eichhorn:2023uge,Howarth:2023bwk,Albandea:2024fui,Abe:2024fpt} (for recent reviews see Refs.~\cite{Finkenrath:2023sjg,Boyle:2024nlh,Finkenrath:2024ptc}).

The adoption of Open Boundary Conditions (OBC) in the Euclidean time direction~\cite{Luscher:2011kk,Luscher:2012av}, instead of the conventional Periodic Boundary Conditions (PBC) is one of the most popular and effective among various strategies proposed to mitigate topological freezing. With OBC, the configuration space of gauge fields becomes simply connected~\cite{Luscher:2011kk}: barriers between topological sectors are removed and the MCMC dynamics of topological modes are now dominated by diffusive phenomena~\cite{McGlynn:2014bxa}, thereby drastically reducing the severity of topological freezing. However, this comes at the price of introducing unwanted boundary effects, as now only field fluctuations sufficiently far from the boundaries are physical, leading to enhanced finite-volume effects. Moreover, translation invariance is lost, hindering for example the proper definition of a global topological charge. In recent years, a method that has been proven to be very effective in circumventing this issue---while at the same time retaining the benefits of OBC simulations---is the \emph{Parallel Tempering on Boundary Conditions} (PTBC) algorithm. After its first introduction in $2d$ $\CP^{N-1}$ models~\cite{Hasenbusch:2017unr} (see also~\cite{Berni:2019bch,Bonanno:2022hmz}), it has been widely employed also in $4d$ gauge theories, both in the pure-gauge case~\cite{Bonanno:2020hht, Bonanno:2022yjr, Bonanno:2023hhp, Bonanno:2024nba, Bonanno:2024ggk} and with dynamical fermions~\cite{Bonanno:2024zyn}. The idea is to perform a tempering on the boundary conditions within a parallel tempering framework by simultaneously simulating several lattices with different boundary conditions, interpolating between OBC and PBC. Such lattice replicas are allowed to swap gauge configurations at equilibrium (i.e., via a standard Metropolis accept/reject step), so that quickly decorrelated fluctuations generated with OBC are transferred to the PBC system, where all observables are computed free of boundary effects.

The present work can be firmly placed within this context, i.e., algorithmic development aimed at alleviating topological freezing in lattice gauge theories. Our goal is to introduce a novel numerical strategy to mitigate this computational problem, combining ideas previously presented in Refs.~\cite{Bonanno:2024udh, Bulgarelli:2024brv}. Although this new proposal shares its basic underlying philosophy with the PTBC algorithm---namely, to combine OBC and PBC to accelerate the MCMC dynamics of topological modes while neutralizing unwanted boundary effects---it is actually rooted on rather different and peculiar ingredients: out-of-equilibrium MCMC simulations~\cite{Caselle:2016wsw, Caselle:2022acb} and flow-based approaches~\cite{Albergo:2019eim,Cranmer:2023xbe}.

At the core of our approach lies a simple and general question: given a field configuration sampled from a starting probability distribution (the \emph{prior}), can it be transformed in a controlled manner, so that it follows a different probability distribution that closely approximates the desired one (the \emph{target})? If the prior distribution features only mild autocorrelations and the transformation itself (the \emph{flow}) is both efficient to find and to sample from, these elements can be combined in a robust strategy to mitigate critical slowing down in lattice gauge theories. The development of the so-called \emph{trivializing map}~\cite{Luscher:2009eq} represented the first major effort in the construction of such a flow transformation, finding however limited success~\cite{Engel:2011re}. More recently, rapid progress in the field of deep learning has provided the tools to construct much more flexible and complex flow transformations, most notably with the implementation of Normalizing Flows (NFs)~\cite{rezende2015,papamakarios2021} for lattice field theory sampling~\cite{Albergo:2019eim, Nicoli:2020njz}. Such architectures possess several desirable features, in particular their expressiveness, allowing them to tackle complicated distributions, and their exactness, as effects due to differences between the inferred and the target distributions can be systematically removed. In recent years, significant progress has been made by the lattice field theory community in the application of different NF architectures to a variety of models, ranging from scalar theories~\cite{Albergo:2019eim, Nicoli:2019gun, Nicoli:2020njz, DelDebbio:2021qwf, Gerdes:2022eve, Singha:2022icw, Chen:2022ytr, Caselle:2023mvh, Albandea:2023wgd, Kreit:2025cos, Schuh:2025gks} to gauge theories~\cite{Kanwar:2020xzo, Boyda:2020hsi, Favoni:2020reg, Bacchio:2022, Singha:2023xxq, Abbott:2023thq, Gerdes:2024rjk}, including formulations with dynamical fermionic variables as well~\cite{Albergo:2021bna, Finkenrath:2022ogg, Albergo:2022qfi, Abbott:2022zhs, Abbott:2024kfc}. 

This new generation of flow-based samplers, however, features its own set of challenges. In particular, finding the optimal NF parameters to flow efficiently from one distribution to another requires a potentially very delicate and expensive training procedure. Concretely, training costs currently suffer from poor scaling, in particular when the number of the relevant degrees of freedom involved in the 
model under study increases (e.g., with larger volumes in units of the lattice spacing), see Refs.~\cite{DelDebbio:2021qwf, Abbott:2022zsh, Komijani:2023fzy, Abbott:2023thq}.

A different flow-based approach built on non-equilibrium MCMC (NE-MCMC) simulations addresses this scaling issue directly. This framework is based on two fundamental results in non-equilibrium statistical mechanics, i.e., Jarzynski's equality~\cite{Jarzynski:1996oqb, Jarzynski:1997ef, Jarzynski:1998ef} and Crooks' theorem~\cite{Crooks:1998,Crooks_1999} and in the last decade it has been successfully applied in lattice field theory. Specifically, its primary application has been the high-precision determination of free energy differences~\cite{Caselle:2016wsw}, in particular for the equation of state~\cite{Caselle:2018kap}, the running coupling~\cite{Francesconi:2020fgi}, the entanglement entropy~\cite{Bulgarelli:2023ofi, Bulgarelli:2024onj} and the Casimir effect~\cite{Bulgarelli:2025riv}. More recently, the same idea has been naturally repurposed as a flow-based approach for the mitigation of critical slowing down~\cite{Bonanno:2024udh, Vadacchino:2024lob}: the current work represents the next step in this direction. A key advantage of the non-equilibrium approach is the well-understood scaling behaviour: in particular, tests in a variety of models show that sampling costs grow linearly with the number of degrees of freedom varied during the flow transformation. 

In their basic implementation, out-of-equilibrium simulations require no training and can achieve efficient sampling with no extra costs. Yet, despite their favorable scaling, they can still require significant amount of computational resources. Interestingly, this purely \textit{stochastic} approach can be naturally combined with the \textit{deterministic} transformations underlying NFs: the resulting architecture, denoted as Stochastic Normalizing Flows (SNFs)~\cite{wu2020stochastic,Caselle:2022acb}, has found natural applications in scalar field theories~\cite{Caselle:2022acb, Caselle:2024ent, Bulgarelli:2024yrz} and, most relevant for this work, in the $\SU(3)$ Yang-Mills theory in 4 spacetime dimensions~\cite{Bulgarelli:2024brv}. SNFs still retain the same desirable scaling properties of NE-MCMC, while at the same time markedly improving its computational efficiency: even more importantly, this is obtained with very limited training costs, a direct consequence of the stochastic nature of these flows. 

It is worth noting that related ideas have appeared in different contexts. NE-MCMC is equivalent to Annealed Importance Sampling~\cite{Neal2001}, which has been reworked recently in the so-called Sequential Monte Carlo~\cite{Dai2022} and also combined with normalizing flows~\cite{arbel2021annealed,Matthews:2022sds}. Recent developments in sampling with Langevin dynamics~\cite{Albergo:2024trn} can be seen as a continuous-time realization of SNFs. Likewise, applications to lattice field theories of diffusion models~\cite{Wang:2023exq, Zhu:2024kiu, Aarts:2024rsl, Zhu:2025pmw, Aarts:2025lpi} also bear several similarities with the ones described in this work: a fundamental difference is that in diffusion models the path between two distributions is defined \textit{implicitly}, whereas the protocol underlying NE-MCMC and SNFs is defined \textit{explicitly}.

In this work, we apply both non-equilibrium Monte Carlo and Stochastic Normalizing Flows as flow-based approaches for efficient topology sampling in the four-dimensional $\SU(3)$ Yang-Mills theory. This model exhibits particularly severe topological freezing near the continuum limit, thus offering an ideal test-bed for flow-based approaches before moving to full QCD simulations. 
In Section~\ref{sec:nemcmc} we introduce the general features of the non-equilibrium Monte Carlo approach and describe our lattice gauge theory setup, including the definition of OBC and of the topological observables of interest. Section~\ref{sec:scaling} presents a careful analysis of the scaling of the sampling costs of non-equilibrium simulations, both from a general perspective and from the point of view of flows connecting OBC with PBC. Section~\ref{sec:snf} focuses on the definition of the customized Stochastic Normalizing Flow used in this work, which has been designed to act specifically on the open boundaries, and on its superior performance with respect to the purely stochastic counterpart. Finally, in Section~\ref{sec:continuum} we discuss the application of this family of flows to simulations at fine lattice spacing, where topological freezing is most severe: here we outline a strategy to sample topological observables towards the continuum limit and present results to further corroborate the effectiveness of this approach. Section~\ref{sec:conclusions} concludes with a broader discussion of future developments, both in terms of advancements in the flow architectures and of applications to more challenging theoretical setups.

\section{Non-equilibrium Monte Carlo simulations in lattice field theory}\label{sec:nemcmc}

On the lattice, given an appropriately discretized action $S[U]$, we wish to compute the vacuum expectation value of a given observable $\mathcal{O}$ as
\begin{equation}
\label{eq:vev}
    \langle \mathcal{O} \rangle_p = \int \dd U  \; \mathcal{O}(U) \, p(U) = \frac{1}{Z} \int \dd U  \; \mathcal{O}(U) \, e^{-S[U]},
\end{equation}
where $p(U) = e^{-S[U]}/Z$ will be referred to as the \textit{target} Boltzmann probability distribution, with
\begin{equation}
\label{eq:partition_function}
    Z_p = \int \dd U  \; e^{-S[U]}
\end{equation}
being the partition function, from which we can immediately define the dimensionless free energy $F = - \log Z$. In a standard Monte Carlo simulation, field configurations $U$ are sampled directly from $p(U)$ by updating them sequentially along a Markov Chain. More precisely, the updating algorithm is characterized by a transition probability $P_p$ which is constructed such that the chain converges to the distribution $p$, called the \textit{equilibrium} distribution. To ensure this, it is standard procedure at the beginning of a simulation to undergo a burn-in period called \textit{thermalization}, after which the Markov chain is assumed to be at equilibrium.

Recent advances in non-equilibrium statistical mechanics, however, enable, under certain conditions and following precise procedures, to perform simulations \textit{out of equilibrium}. In particular, Jarzynski's equality~\cite{Jarzynski:1997ef,Jarzynski:1998ef} represents a fundamental result in this regard: in the case of Markov Chain Monte Carlo (MCMC) simulations, it allows for the calculation of ``equilibrium'' quantities (namely, differences in free energy) from those evaluated on non-thermalized Markov Chains. In particular it is possible to leverage this identity to compute expectation values (in particular in lattice field theory) with a Non-Equilibrium Markov Chain Monte Carlo (NE-MCMC), which we describe in the following. 

In this approach, we build non-equilibrium ``evolutions'' that start from a prior distribution $q_0 = \exp(-S_0)/Z_0$ and reach a target distribution $p = \exp(-S)/Z$, which we aim to sample from. More precisely, each evolution is composed of a \textit{sequence} of $\nstep$ field configurations $U_n$:
\begin{equation}
\label{eq:NE-MCMC_seq}
  \U : \;U_0 \stackrel{P_{\lambda(1)}}{\longrightarrow} \; U_1 \;
  \stackrel{P_{\lambda(2)}}{\longrightarrow} \; U_2 \;
  \stackrel{P_{\lambda(3)}}{\longrightarrow} \; \dots \;
  \stackrel{P_{\lambda(\nstep)}}{\longrightarrow} \; U_{\nstep} \equiv U.
\end{equation}
for which we use the shorthand $\U=[U_0,U_1,\dots,U]$.
At the beginning we have a configuration $U_0$, sampled directly from $q_0$: the latter can be, for example, a Markov Chain at equilibrium or a known analytical distribution one can sample directly from. Then, the evolution proceeds using a composition of Monte Carlo updates with transition probabilities $P_{\lambda(n)}$ (the arrows in the above sequence) which satisfy detailed balance. 

Note that the transition probabilities change throughout the evolution according to a (set of) parameter(s) $\lambda(n)$, called the \textit{protocol}. Each $P_{\lambda(n)}$ is defined with an equilibrium distribution proportional to $\exp(-S_{\lambda(n)})$: the dependence on the protocol $\lambda(n)$ is explicit in the action $S_{\lambda(n)}$, which interpolates (in general completely arbitrarily) between the prior and the target. The only exception is the last transition probability, which is fixed to have the target distribution as equilibrium distribution, i.e., $P_{\lambda(\nstep)} \equiv P_p$ or, equivalently, $\lambda(\nstep)$ must coincide with the value of the target distribution we want to sample from.
The fact that $\lambda(n)$ (and, thus, $P_{\lambda(n)}$) changes after each Monte Carlo update defines a true non-equilibrium evolution: since we do not let it thermalize at each step, the Markov Chain is never at equilibrium.

\begin{figure}
    \centering
    \includegraphics[width=0.75\linewidth]{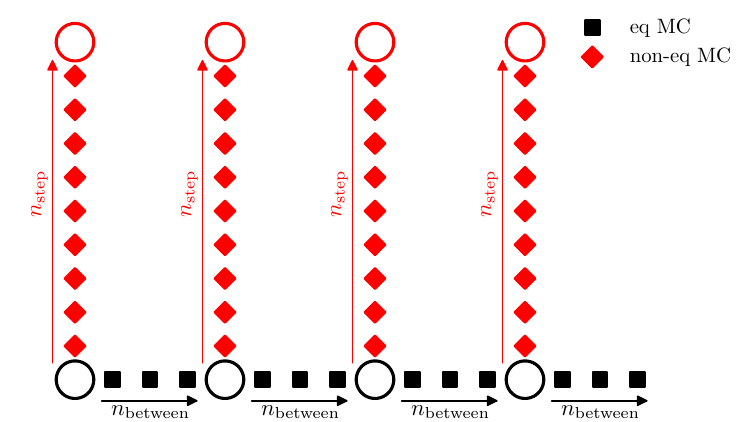}
    \caption{Scheme of a typical non-equilibrium simulation. A thermalized configuration (black circle) is sampled from the prior distribution every $\nbetween$ MCMC steps (black squares); and an out-of-equilibrium evolution starts from it, following a given protocol $\lambda$ for $\nstep$ MCMC steps (red diamonds) until the desired target distribution is reached. The work $W$ of Eq.~\eqref{eq:work} is computed along each evolution, while the value of the desired observable(s) is calculated in the last configuration (red circle). The estimators of Eqs.~\eqref{eq:estimator} and \eqref{eq:jar} are obtained by taking the average $\neavg{\dots}$ across different evolutions.}
    \label{fig:cartoon}
\end{figure}

In order to compute the expectation values of Eq.~\eqref{eq:vev} with NE-MCMC we use the estimator
\begin{equation}
\label{eq:estimator}
    \langle \mathcal{O}(U) \rangle_p  =  \frac{\langle \mathcal{O}(U) \, e^{-W (\U)} \rangle_{\mathrm{f}}}{\langle e^{-W (\U)} \rangle_{\mathrm{f}}}.
\end{equation}
Here, we indicate with $\neavg{\dots}$ an average over all evolutions of the type of Eq.~\eqref{eq:NE-MCMC_seq}; moreover, $W$ is the dimensionless work done on the system during the non-equilibrium transformation from the initial to the final state:
\begin{equation}
\label{eq:work}
W(\U) =\sum_{n=0}^{\nstep-1} \left\{ S_{\lambda(n+1)}\left[U_{n}\right] - S_{\lambda(n)}\left[U_n\right] \right\}.
\end{equation}
Finally, we can write down Jarzynski's equality~\cite{Jarzynski:1997ef, Jarzynski:1998ef}:
\begin{equation}
    \label{eq:jar}
    \langle e^{-W(\U)} \rangle_{\mathrm{f}} = e^{-\Delta F} = \frac{Z_p}{Z_{q_0}},
\end{equation}
which connects the average of the exponential of the work on non-equilibrium evolutions with the difference in free energy between the system described by the target and the prior probability distributions. We show in Fig.~\ref{fig:cartoon} a scheme of a typical NE-MCMC simulation.

\subsection{Some insights on NE-MCMC and its metrics}

Let us first formally define the $\neavg{\dots}$ average of Eq.~\eqref{eq:estimator}.
Once the protocol (i.e., $\lambda(n)$ and $\nstep$) and the Monte Carlo update (i.e., the details of $P_{\lambda(n)}$) have been chosen, we can define the \textit{forward} $\Pf$ and \textit{reverse} $\Pre$ transition probabilities for a given evolution $\U$ as
\begin{equation}
\label{eq:pf}
\Pf[\U] = \prod_{n=1}^{\nstep} P_{\lambda(n)} (U_{n-1} \to U_n) ,
\end{equation}
and
\begin{equation}
\Pre[\U] = \prod_{n=1}^{\nstep} P_{\lambda(n)} (U_{n} \to U_{n-1}).
\end{equation}
Since the Monte Carlo updates must satisfy detailed balance, it is possible to state Crooks’ fluctuation theorem~\cite{Crooks:1998,Crooks_1999}, which for Markov Chains relates the forward and reverse probability densities of a given NE-MCMC evolution to the dissipation of the sequence $\U$:
\begin{equation}
\label{eq:crooks}
\frac{q_0(U_0) \Pf[\U]}{p(U) \Pre [\U]} = \exp (W(\U) - \Delta F);
\end{equation}
this result can be proved rather easily using the properties of Markov Chain transition probabilities that satisfy detailed balance.
It is useful to introduce the pseudo-heat $Q$ exchanged during each transformation:
\begin{equation}
\label{eq:heat}
\begin{split}
Q(\U)&=\sum_{n=1}^{\nstep} \left\{ S_{\lambda(n)}\left[U_{n}\right] - S_{\lambda(n)}\left[U_{n-1}\right] \right\},
\end{split}
\end{equation}
which takes this form following the First Law of Thermodynamics, $S[U]-S_0[U_0] = W(\U) -Q(\U)$.
We have now all the elements to properly define the expectation values over \textit{forward} evolutions as:
\begin{equation}\label{eq:noneqavg} 
\begin{split}
 \langle \dots \rangle_{\U\sim q_0\Pf} \equiv \langle \dots \rangle_{\mathrm{f}}  &= \int \dd U_0 \dots \dd U \; q_0(U_0) \Pf[U_0,\dots,U] \; \dots \\
  &=  \int \dd \U \; q_0(U_0) \Pf[\U] \; \dots, \\
\end{split}
\end{equation}
where the shorthand $\dd \U = \dd U_0 \dots \dd U$ represents the Haar measure over all intermediate configurations. 
Similarly, the expectation values over \textit{reverse} evolutions can be written as:
\begin{equation} 
\begin{split}
 \langle \dots \rangle_{\U\sim p\Pre} \equiv \langle \dots \rangle_{\mathrm{r}}  &= \int \dd U \dots \dd U_0 \; p(U) \Pre[U_0,\dots,U] \; \dots \\
  &=  \int \dd \U \; p(U) \Pre[\U] \; \dots, \\
\end{split}
\end{equation}
Let us now point out that, since the reverse sequences start from configurations $U\sim p$ at the equilibrium, computing expectation values over $p$ and $p\Pre$ is equivalent. Thus, using Crooks' fluctuation theorem:
\begin{equation}
\begin{split}
    \langle \mathcal{O} \rangle_p = &\int \dd U  \; p(U) \mathcal{O}(U) =\int \dd \U \; p(U) \Pre[U, \dots, U_0] \;\mathcal{O}(U) \\
    =&\int \dd \U \; q_0(U_0) \Pf[\U] e^{-(W (\U) -\Delta F)} \mathcal{O}(U) \;,
    \end{split}
\end{equation}
we obtain Eq.~\eqref{eq:estimator}; setting $\mathcal{O} = 1$ leads to Jarzynski's equality~\eqref{eq:jar}.

Crooks' theorem, Eq.~\eqref{eq:crooks}, gives us some precious intuition: evolutions $\U$ that are equally probably going forward and backwards (i.e., more ``reversible'') will feature a work $W$ equal to $\Delta F$. We are however interested in a formal statement about the reversibility of a given protocol (i.e., a choice of the functional form of $\lambda(n)$, $\nstep$ and the MCMC update). To do this, we can take the (reverse) Kullback–Leibler (KL) divergence $\DKL(p_1 \| p_2)$, which measures the degree of similarity of two probability densities $p_1$ and $p_2$. Using the non-equilibrium average defined in Eq.~(\ref{eq:noneqavg}) we can write down the KL divergence between $q_0\Pf$ and $p\Pre$ as
\begin{equation}
\label{eq:kl}
        \DKL(q_0(U_0) \Pf \| p(U) \Pre) = \langle \log \frac{q_0(U_0) \Pf(\U)}{p(U) \Pre(\U)} \rangle_{\mathrm{f}}  \geq 0,
\end{equation}
and using Crooks' theorem this becomes simply
\begin{equation}
        \DKL(q_0(U_0) \Pf \| p(U) \Pre) = \neavg{W(\U)} - \Delta F =\neavg{\Wd (\U)}.
\end{equation}
Here, we defined the \textit{dissipated work} $\Wd(\U)=W(\U) - \Delta F$, which provides a measure of the dissipation of a given thermodynamic out-of-equilibrium transformation (or, equivalently, protocol) between $q_0$ and $p$. If the protocol is reversible, i.e., the KL divergence vanishes, we have $\neavg{\Wd (\U)} = 0$. Furthermore, given the positivity of the KL divergence, we recover the Second Law of Thermodynamics as well, i.e.
\begin{equation}
    \neavg{\Wd (\U)} \geq 0.
\end{equation}

Aside from considerations on the thermodynamic nature of a Markov Chain out of equilibrium, a question arises naturally: why do we care about the KL divergence of Eq.~\eqref{eq:kl}? It is indeed easy to prove that it is an upper bound for another KL divergence:
\begin{equation}
\label{eq:upper_bound}
    \DKL (q \| p) \leq \DKL (q_0 \Pf \| p \Pre),
\end{equation}
with $q$ being the probability distribution of the system at the end of the evolution, which is not analytically accessible in general and we formally define as
\begin{equation}
        q(U) = \int \dd U_0 \dots \dd U_{\nstep-1} \; q_0 (U_0) \Pf (\U).
\end{equation}
Hence, by minimizing the dissipated work $\neavg{\Wd}$, we also necessarily minimize the KL divergence between our target distribution and the one we generate, which is exactly our goal.

Another relevant metric for non-equilibrium protocols is defined looking at the ratio of the \textit{variance} of the two estimators of $\langle \mathcal{O} \rangle$ appearing in Eq.~\eqref{eq:estimator}: the first, sampling directly from $p$ (neglecting autocorrelations); the second, using NE-MCMC. The ratio of the variances of the estimators defines the so-called Effective Sample Size $\ESS$:
\begin{equation}
\label{eq:ess_true}
    \ESS = \frac{\Var(\mathcal{O})_p}{\Var(\mathcal{O})_{\mathrm{NE}}}.
\end{equation}
This is usually approximated (neglecting correlations between $\mathcal{O}$ and $W$) as
\begin{equation}
\label{eq:ess}
    \hESS = \frac{\langle \exp(-W) \rangle_{\mathrm{f}}^2}{\langle \exp(-2W)\rangle_{\mathrm{f}}} = \frac{1}{\langle \exp(-2\Wd) \rangle_{\mathrm{f}}},
\end{equation}
which can be readily computed for any protocol. While not directly relevant for this work, we note that this metric is exactly related to the variance of the estimator of the free energy given by Jarzynski's equality, Eq.~\eqref{eq:jar}:
\begin{equation}
    \Var \left( \exp (-W) \right) = \exp(-2\Delta F) \left (\frac{1}{\hESS} - 1\right).
\end{equation}

\subsection{Lattice setup and topological observables}

As outlined in the introduction, in this investigation we choose the pure-gauge theory as a test-bed for our novel method. We discretize the pure-gauge SU(3) Yang--Mills theory using the standard Wilson plaquette action on an hyper-cubic $L^4$ lattice. Periodic boundary conditions are taken for every link but those living on a small sub-region of the lattice, which in the following will be called \emph{the defect}. In this region, we allow links to experience different boundary conditions, which are changed through discrete out-of-equilibrium steps from open to periodic through a tunable parameter $\lambda(n)$, with $n$ an integer index labeling the out-of-equilibrium step. The path in parameter space connecting OBC to PBC defines, in the case at hand, the out-of-equilibrium protocol introduced in Sec.~\ref{sec:nemcmc}.

The lattice gauge action $S_{\lambda(n)}$ at the $n^{\text{th}}$ out-of-equilibrium step reads:
\beq\label{eq:action_def}
S_{\lambda(n)}[U_n] = - \frac{\beta}{N} \sum_{x, \mu \ne \nu}  K_{\mu\nu}^{(n)}(x)\Re\Tr\left[P_{\mu\nu}^{(n)}(x)\right],
\eeq
where $U_n$ stands for the collection of gauge links at the $n^{\text{th}}$ out-of-equilibrium step. In Eq.~\eqref{eq:action_def} $N=3$ is the number of colors, $\beta=2N/g^2$ is the inverse bare coupling, $P_{\mu\nu}^{(n)}(x)=U^{(n)}_\mu(x)U^{(n)}_\nu(x+\hat{\mu}) {U^{(n)}_\mu}^\dagger(x+\hat{\nu}) {U^{(n)}_\nu}^\dagger(x)$ is the elementary plaquette operator at site $x$ 
on the plane $(\mu,\nu)$ computed on the configuration $U_n$ at the $n^{\rm th}$ out-of-equilibrium step, $a$ is the lattice spacing, and $K_{\mu\nu}^{(n)}(x)$ is a numerical factor used to modify the boundary conditions through the parameter $\lambda(n)$:
\beq
K_{\mu\nu}^{(n)}(x) = K^{(n)}_\mu(x) K^{(n)}_\nu(x+\hat{\mu}) K^{(n)}_\mu(x+\hat{\nu}) K^{(n)}_\nu(x),
\eeq
\beq
\label{eq:bc_par}
K_{\mu}^{(n)}(x) =
\begin{cases}
\lambda(n), \qquad &\mu=0, \quad x_0=L-a, \quad 0 \le x_1,\,x_2,\,x_3 < L_{\dd},\\
\\[-1em]
1,    \qquad &\text{elsewhere},
\end{cases}
\eeq
with $\lambda(n=0)=0$ denoting OBC and $\lambda(n=\nstep)=1$ denoting PBC. The defect is defined as a cube of size $L_\dd$, and it is placed on the time slice $x_0=L-a$ along the temporal direction $\mu=0$, meaning that only plaquettes including temporal links crossing the defect will ``feel'' the modified boundary conditions. We also stress that, for all values of $\lambda$, the Monte Carlo updates of the gauge configurations were performed using the customary 4:1 mixture of over-relaxation (OR)~\cite{Creutz:1987xi} and heat-bath (HB)~\cite{Creutz:1980zw,Kennedy:1985nu} algorithms, implemented according to the Cabibbo--Marinari prescription~\cite{Cabibbo:1982zn}, i.e., updating the 3 $\SU(2)$ subgroups of $\SU(3)$. In the following we will refer to this combination with the shorthand 1HB+4OR.

Given that in this study we are addressing the infamous topological freezing issue, it is natural to focus on the measurement of the topological susceptibility. In this study, as already outlined in the introduction, all computations of physical quantities are performed at the end of the out-of-equilibrium evolution, where PBC are restored all over the lattice. Therefore, from now on we will just assume PBC for all gauge links. Since periodic boundaries preserve translation invariance, we are allowed to consider the total topological charge $Q$ and compute the topological susceptibility via its standard definition:
\beq
\label{eq:chi}
\chi = \frac{\braket{Q^2}}{V}, \qquad V=L^4,
\eeq
where the expectation value in the presence of PBC appearing here is computed out-of-equilibrium using Jarzynski's equality as explained in Sec.~\ref{sec:nemcmc}.

In this study we discretize the topological charge using the simplest parity-odd lattice formulation of the continuum observable $Q=\frac{1}{16\pi^2}\int \dd^4 x \, \Tr [G_{\mu\nu}\tilde{G}_{\mu\nu}]$, defined in terms of the ``clover'' plaquette:
\beq
Q_{\clov} = 
\frac{1}{2^9 \pi^2} 
\sum_{\mu\nu\rho\sigma 
= \pm1}^{\pm4} \varepsilon_{\mu\nu\rho\sigma} 
\Tr\left[ P_{\mu\nu}(x) P_{\rho\sigma}(x) \right],
\eeq
where it is understood that $\varepsilon_{(-\mu)\nu\rho\sigma}=-\varepsilon_{\mu\nu\rho\sigma}$. As it is well known, unlike in the continuum theory, $Q_{\clov}$ is not integer-valued on the lattice, and it is related to the continuum 
topological charge $Q$ via a finite renormalization~\cite{Campostrini:1988cy,Vicari:2008jw}:
$Q_{\clov}=Z_{Q}Q$. The renormalization factor $Z_{Q}(\beta)<1$ tends to $1$ only in the continuum limit $\beta\to\infty$. Moreover, a naive lattice definition of the topological susceptibility $\chi_\clov = \frac{\braket{Q_{\clov}^2}}{V}$, built in terms of $Q_{\clov}$ would receive a divergent additive renormalization term too due to contact terms~\cite{DiVecchia:1981aev,DiVecchia:1981hh, Campostrini:1989dh,DElia:2003zne}, which would eventually overcome the physical signal as $a\to 0$. To deal with these renormalizations, it is customary to resort to smoothing methods, which is by now a widely employed technique~\cite{Alles:1996nm,Alles:1997qe,DelDebbio:2004ns,DelDebbio:2002xa,DElia:2003zne,Lucini:2004yh,Giusti:2007tu,Vicari:2008jw,Panagopoulos:2011rb,Bonati:2013tt,Ce:2015qha,Ce:2016awn,Berkowitz:2015aua,Borsanyi:2015cka,Bonati:2015sqt,Bonati:2016tvi,
Bonati:2018rfg, Bonati:2019kmf, Bonati:2015vqz,
Petreczky:2016vrs,Frison:2016vuc,Borsanyi:2016ksw,Bonati:2018blm,Burger:2018fvb,Chen:2022fid,Athenodorou:2022aay,Bonanno:2023ple,Durr:2025qtq,Butti:2025rnu}. Once $Q_{\clov}$ is computed on smoothened fields, UV contamination at the scale of the lattice spacing is removed, leading to $Z\simeq 1$, and the effects of the contact term vanish. Thus, after smoothing, the lattice definition of $\chi$ is free of multiplicative and additive renormalizations~\cite{DiVecchia:1981aev,DElia:2003zne,Vicari:2008jw}, and will converge towards the correct (finite) continuum limit~\cite{Ce:2015qha,Ce:2016awn}.

On general grounds, smoothing damps UV fluctuations at length scales below a \emph{smoothing radius} $R_s$, while leaving the relevant infrared physics intact (if smoothing is not excessively prolonged). Several smoothing methods have been proposed in the literature, such as
cooling~\cite{Berg:1981nw,Iwasaki:1983bv,Itoh:1984pr,Teper:1985rb,Ilgenfritz:1985dz,Campostrini:1989dh,Alles:2000sc}, stout smearing~\cite{APE:1987ehd,Morningstar:2003gk}, or gradient flow~\cite{Narayanan:2006rf,Luscher:2009eq,Luscher:2010iy,Luscher:2011bx,Lohmayer:2011si}, and they have all been shown to be numerically equivalent when properly matched to each other, i.e., when smearing parameters are chosen so as to correspond to the same value of $R_s$, see Refs.~\cite{Alles:2000sc,Bonati:2014tqa,Alexandrou:2015yba}. In this work, due to its computational inexpensiveness, we adopt cooling, and define our lattice charge and susceptibility as:
\beq
\QL &=& Q_{\clov}[U_{\cool}],\\
\chiL &=& \frac{\braket{\QL^2}}{V},
\eeq
where $U_{\cool}$ denotes the gauge links after applying $n_\cool$ cooling steps. In the case of standard Wilson cooling, the following relation between the number of cooling steps and the smoothing radius has been established by matching with the Wilson gradient flow~\cite{Bonati:2014tqa}:
\beq
\frac{R_s}{a} = \sqrt{\frac{8}{3}n_{\cool}}.
\eeq
The dependence of $R_s$ on $\sqrt{n_{\cool}}$ actually stems from a general feature of smoothing methods: the smoothing radius is always proportional to the square root of the amount of smoothing performed because all smoothing algorithms act as diffusive processes. As an example, using the Wilson flow, $R_s/a=\sqrt{8n_\cool/3} = \sqrt{8t/a^2}$, with $t/a^2$ the flow time, would give an equivalent smoothing radius choosing $t/a^2=n_\cool/3$~\cite{Bonati:2014tqa}. In this work, we adopted $n_{\cool}=60$ for all values of $\beta$ explored, thus corresponding to $R_s \simeq 12.6a\sim 0.4L$ in all cases.

The simulation code used to numerically simulate the lattice setup described so far can be found in the public release~\cite{JARTOP}, based on a modification of~\cite{PTBC}, which is in turn based on a modification of~\cite{yangmills}.

\section{Scaling of NE-MCMC in the boundary conditions}
\label{sec:scaling}

In this section we will discuss the scaling features of a particular non-equilibrium simulation, in which the role of the protocol $\lambda(n)$ appearing in Eq.~\eqref{eq:NE-MCMC_seq} and Eq.~\eqref{eq:pf} is to change the parameter $K_{\mu}^{(n)}(x)$, as already described in Eq.~\eqref{eq:bc_par}. In particular, the evolution in the boundary conditions (from open to periodic) is described by a unique parameter for all the lattice links that are part of the defect. For all the non-equilibrium evolutions performed in this work we used a linear protocol for the parameter $\lambda(n)$; we leave the investigation of more efficient protocols to future work. We thus have just two parameters: the number of degrees of freedom that ``feel'' the defect $\ndof=(L_\dd/a)^3$, and the number of out-of-equilibrium steps $\nstep$. The goal of the present section is to understand how the performances of the NE-MCMC scale with these two quantities.

\begin{figure}[!t]
    \centering
    \includegraphics[width=\textwidth]{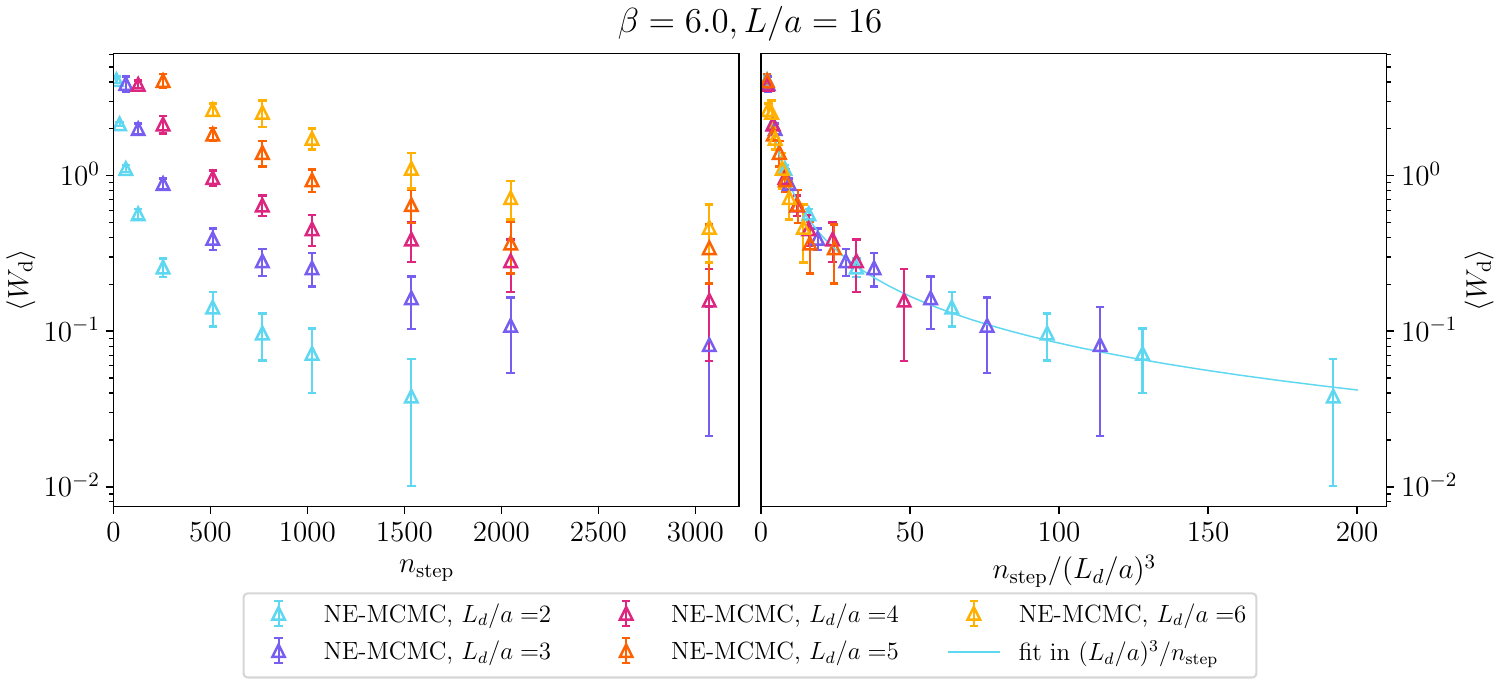}
    \caption{Results for the Kullback-Leibler divergence of Eq.~\eqref{eq:kl} for NE-MCMC in the boundary conditions as a function of the number of steps in the flow $\nstep$ (left panel) and as a function of $\nstep$ divided by the size of the defect (right panel). All results obtained on a $16^4$ lattice at $\beta=6.0$.}
    \label{fig:nemc_6.0_dkl_nstep_scaling}
\end{figure}

\begin{figure}[!t]
    \centering
    \includegraphics[width=\textwidth]{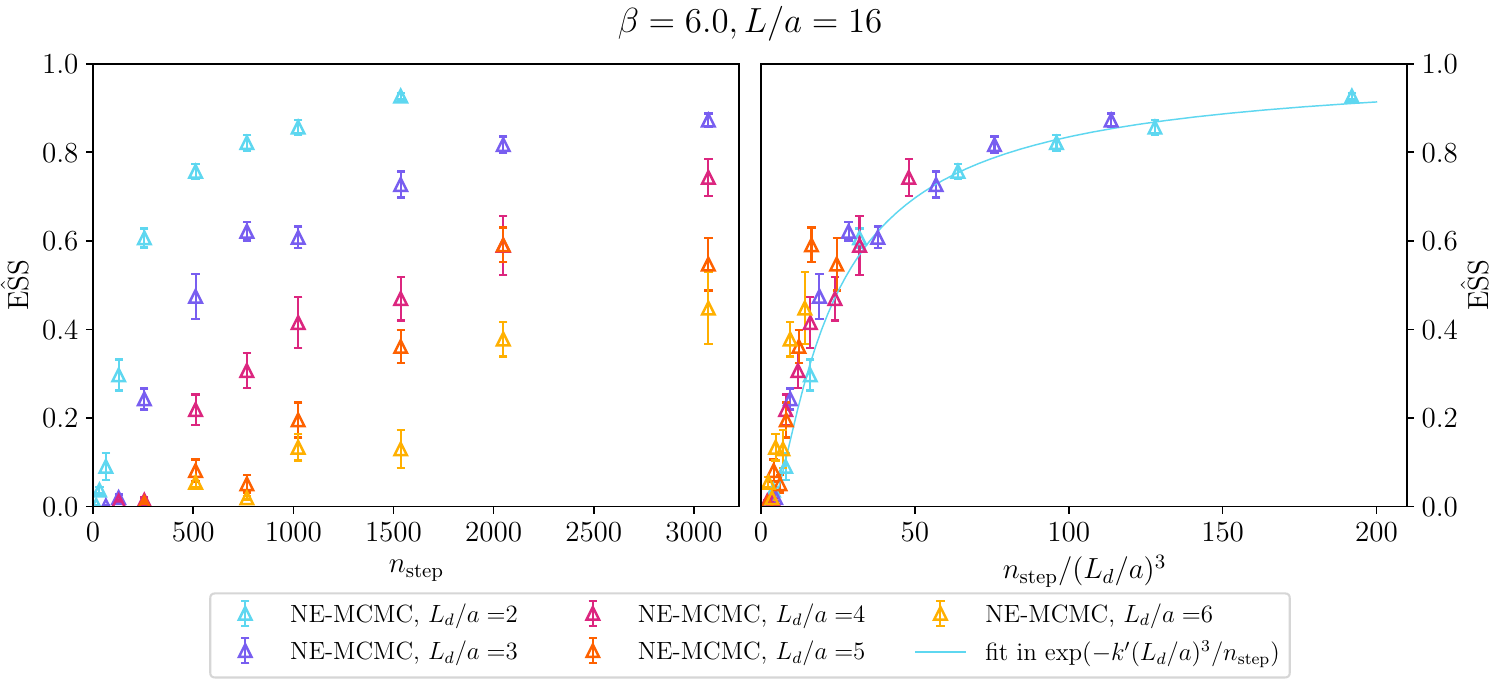}
    \caption{Results for the effective sample size of Eq.~\eqref{eq:ess} for NE-MCMC in the boundary conditions as a function of the number of steps in the flow $\nstep$ (left panel) and as a function of $\nstep$ divided by the size of the defect (right panel). All results obtained on a $16^4$ lattice at $\beta=6.0$.}
    \label{fig:nemc_6.0_ess_nstep_scaling}
\end{figure}

For this purpose, we performed a first batch of NE-MCMC simulations at a relatively coarse spacing at $\beta=6.0$, using cubic defects of different sizes $L_\dd$, ranging from 2 to 6, and varying the duration of the evolution, expressed in units of MCMC updates as $\nstep$ (see Fig.~\ref{fig:cartoon}). We also fix the frequency with which we sample the prior distribution (the one with the OBC defect) to be $\nbetween=5$: this is the number of MCMC updates between starting configurations of subsequent evolutions, see Fig.~\ref{fig:cartoon}. The main aim of this numerical test is to check the behavior of two metrics, the KL divergence of Eq.~\eqref{eq:kl} (equivalent to the dissipated work $\Wd$) and the $\hESS$ of Eq.~\eqref{eq:ess}, as a function of $\nstep$. Results are reported in the left panels of Fig.~\ref{fig:nemc_6.0_dkl_nstep_scaling} and Fig.~\ref{fig:nemc_6.0_ess_nstep_scaling}.

As expected, the value of the dissipated work $\neavg{\Wd}$ decreases rather rapidly when increasing $\nstep$, as the evolutions in the boundary condition parameters are performed more slowly and approach a reversible transformation. This provides a precious upper bound on the similarity between the (analytically intractable) non-equilibrium probability distribution at the end of the evolution and the probability distribution with PBC, see Eq.~\eqref{eq:upper_bound}.
Similarly, the $\hESS$ approaches larger values fairly quickly when $\nstep$ grows; this points at a greatly reduced variance of the weight $\exp( -W)$ and, in good approximation, at a smaller variance of the estimator of Eq.~\eqref{eq:estimator}.

Naturally, the cost in terms of out-of-equilibrium Monte Carlo updates (quantified by $\nstep$) to reach a given target metric (either $\neavg{\Wd}$ or $\hESS$) strongly depends on the size of the defect. This is not surprising, as larger defects naturally induce stronger finite-size effects which in turn require a bigger effort to be removed. The precise scaling relation is easily observed in the right panels of Fig.~\ref{fig:nemc_6.0_dkl_nstep_scaling} and Fig.~\ref{fig:nemc_6.0_ess_nstep_scaling}. Here, the dissipated work and the $\hESS$ are once again plotted, but this time as a function of $\nstep/(L_\dd/a)^3$, that is, the duration of the evolution in terms of MCMC steps divided by the spatial volume of the defect in lattice units. The latter quantity is exactly the number of degrees of freedom modified along the evolution itself. 
Data for different defect sizes collapse neatly on the same curve, which represents precisely the scaling function of NE-MCMC metrics: for example, fixing the defect size $L_\dd$, the value of $\nstep$ needed to reach a target metric (e.g., $\hESS=0.4$) can be immediately derived just looking at these results. 

\subsection{Understanding the scaling with the degrees of freedom}

It is worthwhile to understand the NE-MCMC scaling a bit more precisely: in particular, the
reason why $\neavg{\Wd}$ seems to depend uniquely on $\nstep/\ndof$, with $\ndof$ being the number of degrees of freedom that are varied throughout a non-equilibrium trajectory. This fact is far from being limited to evolutions from OBC to PBC (where it was already observed in $2d$ $\CP^{N-1}$ model~\cite{Bonanno:2024udh}). It is also present, for instance, when changing the inverse coupling $\beta$ in $\SU(3)$ pure gauge theory~\cite{Bulgarelli:2024brv} and when exchanging slabs between lattices in O$(N)$ spin models~\cite{Bulgarelli:2025riv}.

Let us look at the dissipated work once again, writing it as
\begin{equation}
\label{eq:wd_scaling1}
 \neavg{\Wd} = \neavg{W} - \Delta F \simeq \delta \lambda \sum_{n=0}^{\nstep-1} \left\{  \neavgn{\pd{S_{\lambda(n)}}{\lambda} } - \eqavg{\pd{S_{\lambda(n)}}{\lambda} } \right\},
\end{equation}
where we used $1 /\nstep \equiv \lambda_{n+1} - \lambda_{n}$, i.e., assuming for simplicity a linear change in a protocol parameter and that $\lambda(0) = 0$ and $\lambda(\nstep)=1$. We also (approximately) calculated the free energy difference using a basic implementation of the integral method: indeed, the $\eqavg{\dots}$ average is the standard expectation value at equilibrium with respect to the probability distribution defined with $S_{\lambda(n)}$. This is in general very different from the $\neavgn{\dots}$ average, which it is calculated during a non-equilibrium evolution for a specific protocol, and also depends strongly on the details of the latter.

Let us look at the derivative of the action with respect to the protocol parameter $\lambda$ first: for evolutions in the boundary conditions and specifying the action to be Eq.~\eqref{eq:action_def}, this term is simply the sum of the plaquettes that touch the defect\footnote{In the case of evolutions in $\beta$, the $\partial S_{\lambda(n)} / \partial \lambda$ term would be the sum of all the plaquettes on the lattice.}.
In this case we can write it down approximately as
\begin{equation}
    \pd{S_{\lambda(n)}}{\lambda} \simeq - 6 \beta \, \ndof \, \lambda(n) \, P_d ,
\end{equation}
with $P_d$ being the average of the plaquettes that contain two of the defect links indicated in Eq.~\eqref{eq:bc_par} and $\ndof = (L_d/a)^3$; as we are interested only in a qualitative behavior, we ignore the plaquettes containing only one defect link. Now the dissipated work becomes
\begin{equation}
 \neavg{\Wd} \simeq 6\frac{\ndof}{\nstep} \beta \sum_{n=0}^{\nstep-1} \lambda(n) \left(\eqavg{P_d} - \neavgn{P_d} \right).
\end{equation}
The question is, what is the behavior of the defect plaquettes at each step $n$ of a non-equilibrium evolution with respect to its corresponding value at equilibrium (i.e., for the same parameter $\lambda(n)$)? Intuition suggests that it should vanish when going towards equilibrium, i.e., in the limit $\nstep \to \infty$. Thus, we assume\footnote{This is not arbitrary, as linear response theory generally predicts this behavior at first order~\cite{Sivak_2012}.} that this quantity is directly proportional to the speed of the evolution:
\begin{equation}
    \eqavg{P_d} - \neavgn{P_d} \sim \frac{1}{\nstep}.
\end{equation}
Recalling that in our case $\lambda(n) = n/\nstep$, this finally gives us a very simple qualitative behavior for the average dissipated work:
\begin{equation}
\label{eq:wd_appr}
 \neavg{\Wd} \sim 6\frac{\ndof}{\nstep} \beta \sum_{n=0}^{\nstep-1} \frac{n}{\nstep^2} \times K(\beta) \sim 3 \beta \frac{\ndof}{\nstep} \times K(\beta),
\end{equation}
where we the $K$ factor contains a residual dependence on $\beta$: a proper investigation of a good approximation is left to future work.
Finally, we have recovered an explicit dependence of the dissipated work (or reverse KL divergence) $\neavg{\Wd}$ on the ratio $\nstep/\ndof$: this is further supported by the excellent fit of the data of Fig.~\ref{fig:nemc_6.0_dkl_nstep_scaling} with a $1/(\nstep/\ndof)$ behavior. 

\begin{figure}[t]
    \centering
    \includegraphics[width=0.75\linewidth]{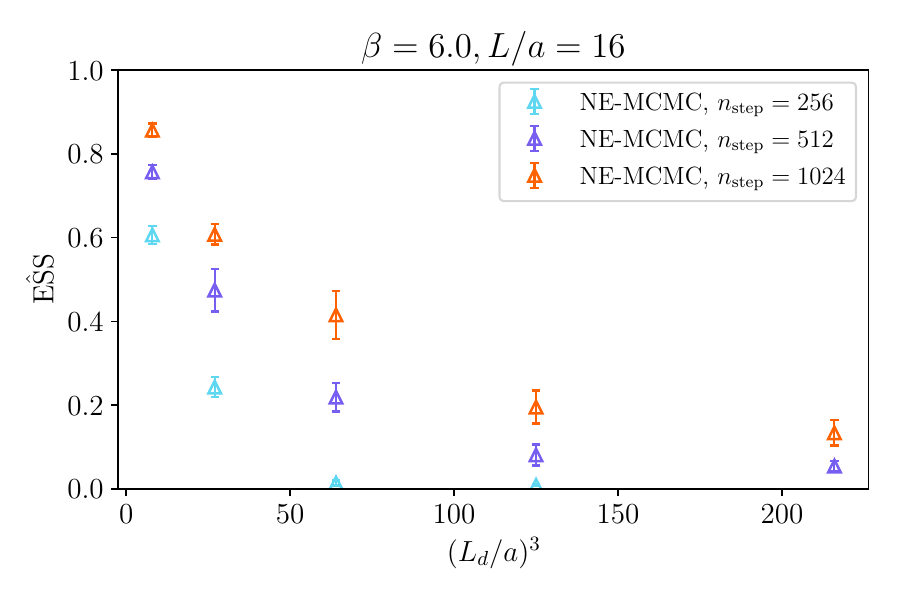}
    \caption{Effective Sample Size $\hESS$ as a function of the number of links on the defect, for three fixed NE-MCMC architectures.}
    \label{fig:ess_d}
\end{figure}

Another useful analysis can be made on the Effective Sample Size: it has been observed in the past~\cite{Abbott:2022zsh,Finkenrath:2022ogg} that for a \textit{fixed} flow architecture, the $\hESS$ decreases exponentially with the number of degrees of freedom in the system\footnote{This statement is usually expressed as $\hESS(\ndof) = \ESS(\ndof^{(0)})^{\ndof/\ndof^{(0)}}$.}, i.e.
\begin{equation}
    \hESS(\ndof) = \exp(-k \, \ndof),
\end{equation}
which is observed also in the case of NE-MCMC flows for fixed $\nstep$, see Fig.~\ref{fig:ess_d}.

However, we also observe from Fig.~\ref{fig:nemc_6.0_ess_nstep_scaling} that the $\hESS$ is, to a very good approximation, a function of $\nstep/\ndof$ and not just of $\ndof$; hence, it is natural to write that
\begin{equation}
\label{eq:ess_fit}
    \hESS(\ndof) = \exp\left(-k' \, \frac{\ndof}{\nstep}\right),
\end{equation}
and this is confirmed by the excellent qualitative agreement when fitting the data points in Fig.~\ref{fig:nemc_6.0_ess_nstep_scaling}. This is a striking example of how incorporating non-equilibrium MC updates in a flow-based approach (as in the case of NE-MCMC) naturally provides an \textit{exponential} improvement with respect to a given fixed architecture. This analysis strongly suggests that NE-MCMC (and related approaches) offer a compelling framework to tackle the issue of scaling flow-based samplers to problems characterized by large $\ndof$.

\section{Accelerating NE-MCMC with Stochastic Normalizing Flows}\label{sec:snf}

The estimator of Eq.~\eqref{eq:estimator} is unbiased: however, since it relies on an exponential average, it has to be handled with care. In particular, it can suffer from high variance (i.e., low $\hESS$) when the dissipated work is large: sampling $p$ in such conditions would require extremely large statistics to be reliable, as the average on the evolutions strongly depends on a few \textit{rare events} that populate the tail of the distribution. For examples of the distribution of the weight $\exp (-W)$ see Refs.~\cite{Bonanno:2024udh,Bulgarelli:2024brv}; moreover, note that the number of samples (i.e., evolutions) that are needed grows exponentially with $\neavg{\Wd}$, see Ref.~\cite{Jarzynski_2006} for an in-depth discussion.

Naturally, this problem can be solved by increasing $\nstep$: in the asymptotic limit, the transformations become quasi-static, meaning that the system remains close to equilibrium. In such conditions, $\langle \Wd \rangle$ is small and $\exp (-W)$ fluctuates mildly: the exponential average is under control. While effective, this simple strategy can be too expensive for practical purposes, especially when the number of degrees of freedom that one needs to vary in a transformation becomes large.

A more general strategy that can help mitigate the growth of $\Wd$ (and thus, the total cost of the algorithm) is to enhance NE-MCMC with a class of deep generative models called Normalizing Flows (NFs)~\cite{rezende2015}. 
The idea behind a generic NF $g_\rho$ is very simple: it is a diffeomorphism dependent on a set of parameters $\{ \rho \}$ that acts on a configuration $U_0$ sampled from a distribution $q_0$ and transforms it in a different configuration $U = g_\rho(U_0)$ which follows a variational density $q$. Simply using the change of variables theorem we can write it as 
\begin{equation}
    q(U) = q_0(g^{-1}_\rho(U))|\det J_{g_\rho}|^{-1},
\end{equation}
with $\det J_{g_\rho}$ being the determinant of the Jacobian of the NF. 
The power of this approach depends on the fact that a generic NF is built as a composition of $l$ intermediate functions $g_{\rho(n)}$, the so-called coupling layers:
\begin{equation}
    g_\rho (U_0) = g_{\rho(l)} \cdot g_{\rho(l-1)} \cdots g_{\rho(1)} (U_0),
\end{equation}
each of them depending on a subset of parameters $\{\rho(l) \}$. Each of them transforms the field configuration through appropriate masking patterns to guarantee invertibility and an easy calculation of the Jacobian: inside each $g_{\rho(l)}$, neural networks can be employed to increase the expressivity of the full transformation.

In this work, we do not employ NF alone, but use a relatively straightforward implementation of their coupling layers to \textit{assist} NE-MCMC protocols. More precisely, we interleave NE-MCMC updates with NF layers to create a sequence, as follows:
\begin{equation}
\label{eq:SNF_sequence}
   U_0 \stackrel{g_{\rho(1)}}{\longrightarrow} \; g_{\rho(1)}(U_0) \;
  \stackrel{P_{\lambda(1)}}{\longrightarrow} \; U_1 \;
  \stackrel{g_{\rho(2)}}{\longrightarrow} \; g_{\rho(2)}(U_1) \;
  \stackrel{P_{\lambda(2)}}{\longrightarrow} \; \dots \;
  \stackrel{P_{\lambda(\nstep)}}{\longrightarrow} \; U_{\nstep}.
\end{equation} 
This defines a particular instance of Stochastic Normalizing Flows (SNFs)~\cite{wu2020stochastic}, in which every non-equilibrium update $P_{\lambda(n)}$ is preceded by a deterministic transformation $g_{\rho(n)}$. 

A given target distribution $p$ can be sampled with the enhanced protocol of Eq.~\eqref{eq:SNF_sequence} using the same framework of NE-MCMC. In particular, the estimator of Eq.~\eqref{eq:estimator} and the metrics of Eq.~\eqref{eq:kl} and Eq.~\eqref{eq:ess} can be readily employed with one single modification: we now have to use the \textit{variational} work, which for the SNF of Eq.~\eqref{eq:SNF_sequence} can be computed as~\cite{Vaikuntanathan_2011,wu2020stochastic}: 
\begin{align}
\label{eq:work_snf}
    W^{(\rho)}(\U) &= S[U] - S_0[U_0] - Q(\U) -\log J(\U) \\
          &=\sum_{n=0}^{\nstep-1} S_{\lambda(n+1)} \left[g_{\rho(n+1)} (U_n) \right] - S_{\lambda(n)}\left[U_n \right]  -\log J_{g_{\rho(n+1)}}[U_n].
\end{align}
The additional term:
\begin{equation}
\log J(\U) = \sum_{n=0}^{ \nstep-1 } \log J_{g_{\rho(n+1)}}(U_{n}),
\end{equation}
represents the cumulative contribution from the logarithms of the Jacobian determinants, accounting for the change in density induced by the NF layers.

Naturally, for the transformations $g_{\rho(n)}$ to be useful, the parameters $\{ \rho(n) \}$ have to be \textit{trained}, i.e., tuned according to some minimization \textit{training} procedure. In this framework we optimize them by minimizing the Kullback-Leibler divergence of Eq.~\eqref{eq:kl}:
\begin{equation}
\label{eq:min_kl}
\textrm{min}_{ \{ \rho \} } \DKL( q_0 \Pf \| p \Pre) = \textrm{min}_{ \{\rho \}} \langle \Wd^{(\rho)} (\U ) \rangle_\mathrm{f},
\end{equation}
The interpretation is straightforward: SNF parameters are tuned to bring a given protocol as close as possible to equilibrium.

The design of coupling layers is typically guided by encoding the relevant symmetries of the theory directly into the machine learning model. This approach is expected to improve the efficiency of the model and accelerate training~\cite{Taco:2016equiv}. A common strategy for incorporating symmetries into flow-based samplers is to construct equivariant coupling layers~\cite{Kanwar:2020xzo, Abbott:2023thq, kohler2019equivariant}, ensuring that the transformation $g$ commutes with the symmetry.
In our implementation, we use gauge-covariant coupling layers~\cite{Nagai:2021bhh}, where the diffeomorphisms $g_{\rho(n)}$ are essentially stout smearing transformations~\cite{Morningstar:2003gk}; in this work we follow the same straightforward implementation used in Ref.~\cite{Bulgarelli:2024brv} for flows in $\beta$.
The field transformation for a given link is defined as:
\begin{equation}
    U'_\mu (x) = g_{\rho(n)} (U_\mu (x)) = \exp \left(i Q_\mu^{(n)} (x) \right) \, U_\mu (x),
    \label{eq:cl_stout_smearing}
\end{equation}
with $Q_\mu$ Hermitian and traceless: 
\begin{equation}
\begin{split}
     Q_\mu^{(n)} (x) =  \frac{i}{2} \left( (\Omega^{(n)}_\mu (x))^\dagger - \Omega^{(n)}_\mu (x) \right) + \\
     - \frac{i}{2N} \Tr \left( (\Omega^{(n)}_\mu (x))^\dagger - \Omega^{(n)}_\mu (x) \right) ,
\end{split}
\end{equation}
and where $\Omega_\mu^{(n)} (x)$ is a sum of untraced loops based on $x$. 
We have
\begin{equation}
\label{eq:cl_omega}
    \Omega_\mu^{(n)} (x) = C_\mu^{(n)} (x) U^\dagger_\mu (x),
\end{equation}
that is made by a weighted sum over staples:
\begin{equation}
\begin{split}
\label{eq:cl_staples}
    C_\mu^{(n)} (x) =  \sum_{\nu \neq \mu} \rho^+_{\mu \nu}(n,x)  U_\nu(x) U_\mu(x+\hat{\nu}) U_\nu^\dagger(x + \hat{\mu}) \\
     + \rho^-_{\mu \nu}(n,x) U^\dagger_\nu(x-\hat{\nu})U_\mu(x-\hat{\nu})U_\nu(x-\hat{\nu}+\hat{\mu}).
\end{split}
\end{equation}
The coefficients $\rho^{\pm}_{\mu \nu}(n,x)$ represent the parameters tuned in the training procedure; here we take the most general form, in which they depend also whether the staple is in the $+\hat\nu$ or $-\hat\nu$ direction.
These layers can be generalized to work on larger loops, as described in Ref.~\cite{Abbott:2023thq}. 
To ensure invertibility, it is crucial to apply a proper masking procedure: in this case, we apply an even-odd decomposition and then use the transformation of Eq.~\eqref{eq:cl_stout_smearing} one direction at a time, so that each layer $g_{\rho(n)}$ contains 8 different transformations. In this pattern, the links in $C_\mu^{(n)} (x)$ can be considered ``frozen'' while the $U_\mu(x)$ in Eqs.~\eqref{eq:cl_stout_smearing} and \eqref{eq:cl_omega} are the ``active'' ones. 

\subsection{Coupling layers for a defect}

Directly encoding symmetries into the variational \textit{Ansatz} of the coupling layer is not the only strategy to enhance its effectiveness. In the present context, the geometry of the problem itself suggests a design for a deterministic transformation. Ref.~\cite{Bulgarelli:2024yrz} introduced, for the first time, the concept of a \textit{defect coupling layer}, defined as a standard coupling layer restricted to a localized region of interest. Specifically, the layer acts only on a subset of the lattice degrees of freedom, here, the gauge fields, located near the defect, and is conditioned on a fixed set of degrees of freedom, also limited to a localized region of the lattice. As a result, the majority of the lattice remains untouched by the transformation: these degrees of freedom are neither used as inputs nor altered by the coupling layer.

This approach, which \textit{a priori} selects the relevant region of the lattice where the transformation is applied, has been shown to be effective and to drastically reduce the cost of the training compared to the standard approach, where the whole lattice undergoes a transformation~\cite{Bulgarelli:2024yrz}. At first glance, however, this approach might appear to be problematic if one would like the defect to have a global effect on the system, as most of the d.o.f. remain unaffected by the deterministic transformation; this is expected to be the case for defects related to boundary conditions. Nonetheless, it is important to emphasize that the defect coupling layer constitutes only one component of the SNF; the other essential ingredient is the Monte Carlo update, which, in the present framework, always acts globally on the full lattice. This has the effect of spreading the information on the modified defect far from the region where the coupling layer is acting, while the coupling layers accelerate the removal of the effect of OBC in the proximity of the defect.

In practice, in this work we apply the stout smearing transformation defined in Eq.~\eqref{eq:cl_stout_smearing} uniquely in two cases:
\begin{itemize}
    \item on links $U_\mu(x)$ on the defect, i.e., $\hat\mu = \hat0$, $x_0=L-a$ and  $0 \le x_1,\,x_2,\,x_3 < L_{\dd}$,
    \item on links $U_\mu(x)$ which are not themselves on the defect, but for which the corresponding sum of staples appearing in Eq.~\eqref{eq:cl_staples} contains at least a link on the defect.
\end{itemize} 
More precisely, we set the parameters $\rho^{\pm}_{\mu \nu}(n,x)$ to be non-vanishing only if the link being transformed is on the defect, or the corresponding staple has one link on the defect. At this stage we opt not to use any notion of symmetry from the cubic geometry of the OBC defect and we leave all parameters independent. The number of parameters per coupling layer (counting one layer as the composition of the eight masks) grows proportionally with the defect volume $(L_d/a)^3$ and we report it in Table~\ref{tab:nrho}. 

\begin{table}[!t]
    \centering
    \begin{tabular}{|c|c|c|c|c|c|}
    \hline
      $L_d/a$  & 2 & 3 & 4 & 5 & 6\\
         \hline 
       $n_\rho$ & 144 & 432 & 960 & 1800 & 3024 \\
         \hline
    \end{tabular}
    \caption{Number $n_\rho$ of stout smearing parameters appearing in a single defect coupling layer for different defect sizes $L_d/a$.}
    \label{tab:nrho}
\end{table}

We remark here that in the following, an SNF architecture with $\nstep$ steps indicates a combination of one defect coupling layer (i.e., a stout smearing transformation of the relevant links) plus one full update of the whole lattice with the standard 1HB+4OR updates (following the $\lambda(n)$ protocol), repeated $\nstep$ times. This implementation of SNFs is available as a CPU code~\cite{JARTOP} uniquely for sampling and as a PyTorch code~\cite{SNFSU3} (also for GPUs) for both training and sampling.

We performed several training procedures minimizing the KL divergence, i.e., the generalized dissipated work, as in Eq.~\eqref{eq:min_kl}. We chose again a $L/a=16$ hypercubic lattice at $\beta=6.0$ with a defect size in the range $L_d/a \in [2,6]$. We trained SNFs with $\nstep=8$ and $\nstep=16$, but performing the backpropagation separately for each layer: in practice, we minimize the terms in the sum in Eq.~\eqref{eq:min_kl} one by one; see Ref.~\cite{Bulgarelli:2024brv} for a discussion of this procedure and its connection to the work of Ref.~\cite{Matthews:2022sds}. We performed the training for about 1000--2000 iterations using the \texttt{Adam} optimizer~\cite{Kingma:2014vow}, after which the loss for all values of $\nstep$ and $L_d/a$ reaches a plateau.
One of the advantages of this procedure is that the memory consumption during training is independent of $\nstep$; still, the minimization procedure becomes not just more expensive with $\nstep$, but also more difficult.
To overcome this issue, we generalize the methodology employed in Ref.~\cite{Bulgarelli:2024brv} to the case of flows in the boundary conditions. In particular:
\begin{itemize}
    \item from the results of the trainings for a fixed value of $\nstep$ we identify 9 classes of parameters (just 5 in the $L_d/a=2$ case) characterized by the geometry of the cubic defect;
    \item for each class, we take the average $\rho^{\rm (class)}(n)$ of all the corresponding parameters and we multiply it by $\nstep$;
    \item finally, we perform a spline interpolation of $\rho^{\rm (class)}(n) \times \nstep$ in $n/\nstep \in [0,1]$.
\end{itemize}
The spline function is the true result of the training: indeed, we use it to extract the corresponding value of $\rho(n)$ for any value of $\nstep$. In a sense, it can be considered as a peculiar case of transfer learning: we train uniquely in a simple setting (i.e., an SNF with few layers), recognize a particular pattern in the weights, and extrapolate the result for any $\nstep$. We provide more details of the interpolation procedure in Appendix~\ref{sec:app_rho_interpolation}.

\begin{figure}[!t]
    \centering
    \includegraphics[width=\linewidth]{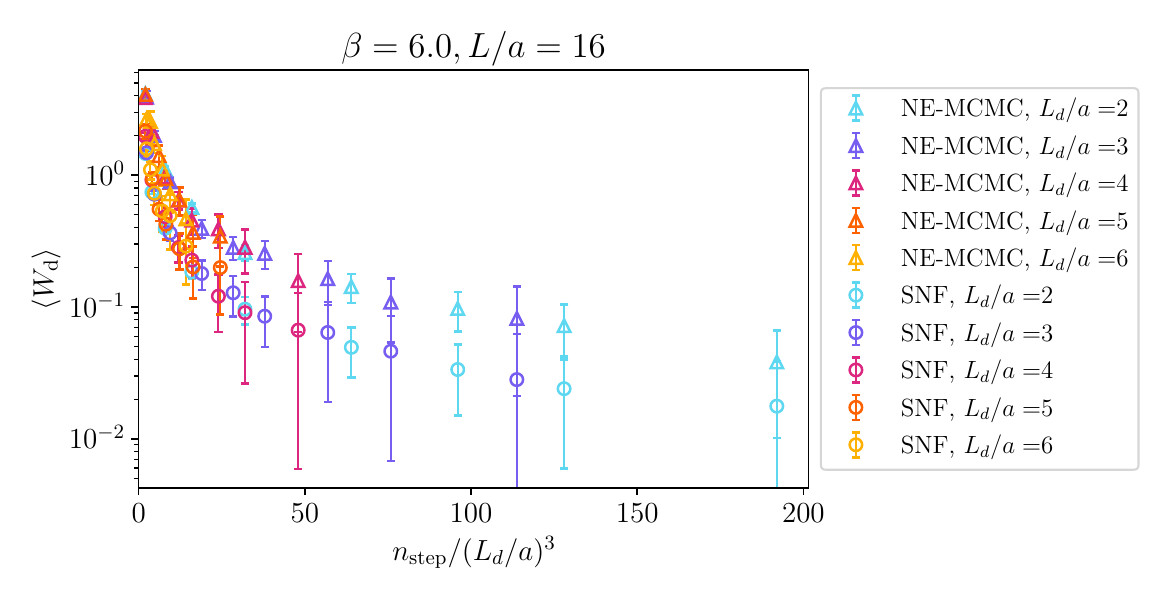}
    \caption{Results for the Kullback-Leibler divergence of Eq.~\eqref{eq:kl} for different flows in the boundary conditions as a function of the number of steps in the flow divided by the volume of the defect. Both NE-MCMC (circles) and SNFs with defect coupling layers (squares) are shown. All results obtained on a $16^4$ lattice at $\beta=6.0$.}
    \label{fig:6.0_dkl_nemc_snf}
\end{figure}

\begin{figure}[!t]
    \centering
    \includegraphics[width=\linewidth]{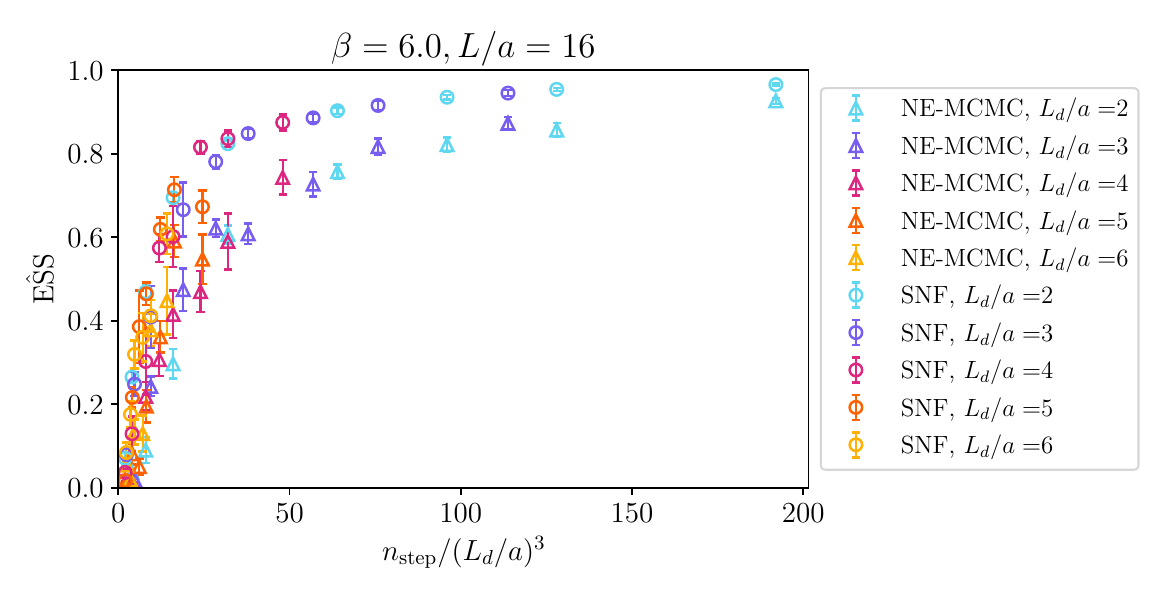}
    \caption{Results for the effective sample size of Eq.~\eqref{eq:ess} for different flows in the boundary conditions as a function of the number of steps in the flow divided by the volume of the defect. Both NE-MCMC (circles) and SNFs with defect coupling layers (squares) are shown. All results obtained on a $16^4$ lattice at $\beta=6.0$.}
    \label{fig:6.0_ess_nemc_snf}
\end{figure}

Finally, we can use the SNFs trained in this way and compare their performances with those of standard NE-MCMC when sampling the target density with PBC. We report in Figs.~\ref{fig:6.0_dkl_nemc_snf} and \ref{fig:6.0_ess_nemc_snf} the comparison between SNFs and NE-MCMC in terms of dissipated work and $\hESS$ for the same combinations of defect sizes and $\nstep$ previously analyzed in Sec.~\ref{sec:nemcmc}. 

The results show a very clear advantage in using SNFs over standard NE-MCMC: the KL divergence drops faster towards zero, indicating more reversible evolutions, while the $\hESS$ grows quicker with $\nstep$, which implies a smaller variance for the estimator of Eq.~\eqref{eq:estimator}. The comparison is fair, as for fixed $\nstep$ the computational effort for each evolution is essentially the same for the two flow architectures. Indeed, the cost of performing the stout smearing transformations around a three-dimensional object is negligible with respect to a full 1HB+4OR update of the much larger four-dimensional lattice.

Training costs in flow-based approaches must be closely monitored, as they generally represent a significant fraction of the total simulations costs when using for example NFs. However, this is not the case for SNFs, as the presence of a fixed protocol makes the training much shorter and, to some extent, almost trivial. First of all, from our tests the training procedure is essentially independent of $\nstep$, so we perform it only for the smallest $\nstep=8$ and $16$ and simply extrapolate for larger $\nstep$ (i.e., slower evolutions); see Appendix~\ref{sec:app_rho_interpolation} for more details. 
Furthermore, these trainings lasted around $10^3$ iterations of the minimization process: to put it in perspective note that, including backpropagation, the cost of an iteration is roughly comparable with generating one sample (i.e., one evolution). 
For the actual simulation with a trained flow we generated $\nev \simeq 10^4$ evolutions for the $\nstep = 16$ case and similar total effort for larger values of $\nstep$: thus, the total cost of the training for each value of $L_d$ is a fraction of the cost of generating enough samples (evolutions) to carefully assessing the metrics of the flow itself.\footnote{The cost of training can become more significant if more complex coupling layers are employed; in that case, an overall cost function that takes into account the computational effort to reach a given $\hESS$ during training would be needed.}.

Such results can be interpreted in essentially two ways. In the first, we keep both the computational cost (i.e., $\nstep$) and the size of the problem (removing the effect of a $(L_d/a)^3$ OBC defect) fixed: doing so, SNFs provide an overall better estimator in any case as the $\hESS$ is always markedly higher. In some cases, one can sample with SNFs where it would be essentially impossible with NE-MCMC. 
The second way to interpret these results is to keep both the size of the problem (i.e., the value of $(L_d/a)^3$) and the quality of the estimator (e.g., the $\hESS$) fixed: one can then ask, what is the relative effort required to reach the value of a certain metric. For SNFs, this appears to be consistently \textit{one third} of the effort required by NE-MCMC: indeed, the curve drawn by the SNF results in Figs.~\ref{fig:6.0_dkl_nemc_snf} and \ref{fig:6.0_ess_nemc_snf} is the same as the purely stochastic one, but compressed horizontally by a factor 3.

We have established the superiority of a rather simple SNF architecture in removing the effects of OBC and in sampling a target distribution with PBC in an unbiased and scalable fashion. However, each value of $\nstep$ defines a different estimator, with a different variance approximated by the corresponding value of the $\hESS$ from Fig.~\ref{fig:6.0_ess_nemc_snf}: which is then the most efficient one? Equivalently, the question is whether it is better to ``spend'' less (in terms of MCMC updates) and be content with a relatively small value of $\hESS$, or to spend more for an estimator with a smaller variance.

\begin{figure}[!t]
    \centering
    \includegraphics[width=0.75\linewidth]{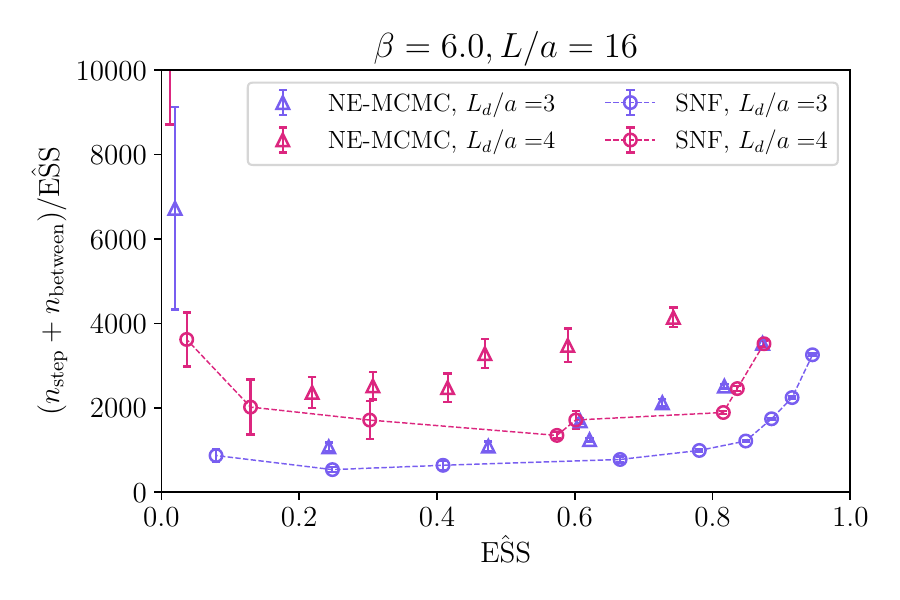}
    \caption{Results for the efficiency factor $(\nstep+\nbetween)/\hESS$ from Eq.~\eqref{eq:eff} for different flows in the boundary conditions as a function of the effective sample size. All results obtained on a $16^4$ lattice at $\beta=6.0$.}
    \label{fig:6.0_eff}
\end{figure}

A cost function $C_{\mathrm{f}}$ to generate $\nev$ samples of an observable $\mathcal{O}$ with the flow-based approaches studied in this work can be written (neglecting autocorrelations in the data and training costs) as the number of evolutions times the cost of a single evolution:
\begin{align}
    C_{\mathrm f} (\nev) &= \nev \times \text{cost}_{\rm ev} = \, \nev \times (\nstep + \nbetween) \\
    & = \frac{\Var_{\mathrm{f}} \, ( \mathcal{O} )}{\text{err} (\mathcal{O})^2} \times (\nstep + \nbetween) = \frac{\Eff_{\mathrm{f}} (\mathcal{O}) }{\text{err} (\mathcal{O})^2},
\end{align}
here we identified the \textit{effective} cost of sampling an observable $\mathcal{O}$ with 
\begin{equation}
    \Eff_{\mathrm{f}} ( \mathcal{O} ) \equiv \Var_{\mathrm{f}} ( \mathcal{O} ) \times (\nstep + \nbetween).
\end{equation}
We can rewrite it as
\begin{equation}
\label{eq:eff}
    \Eff_{\mathrm{f}} ( \mathcal{O} ) \simeq \Var ( \mathcal{O} )\frac{\nstep + \nbetween}{\hESS},
\end{equation}
where we used the definition of $\hESS$ from Eq.~\eqref{eq:ess}. The efficiency of the flow then depends on the variance of $\mathcal{O}$ (a theoretical value that is fixed for a given target distribution $p$) and the ratio $(\nstep+\nbetween)/\hESS$\footnote{At this stage, the role of $\nbetween$ is secondary: since we are effectively neglecting autocorrelations in our samples, increasing the spacing between evolutions has no direct influence in the efficiency of the flow.}, which we show in Fig.~\ref{fig:6.0_eff} as a function of the corresponding $\hESS$. 

First of all, SNFs are consistently more efficient than the corresponding NE-MCMC counterpart, and larger defects are also less efficient as we are ignoring autocorrelations for the moment. Furthermore, flows characterized by small values of $\hESS$ are unsurprisingly very expensive and should be avoided; more interestingly, the largest values of $\hESS$ do not appear to be particularly efficient either. Indeed, there seems to be a typical value of $\hESS$ (or, equivalently, of $\nstep/(L_d/a)^3$) above which it is not worth to increase the quality of the flow, as it becomes too costly. 

It is interesting to analyze the most efficient value of $\nstep/(L_d/a)^3$ (or equivalently, the  ``best'' value of the $\hESS$) using the parametrization of Eq.~\eqref{eq:ess_fit}: in practice, completely neglecting $\nbetween$, we wish to minimize
\begin{equation}
\label{eq:eff2}
    \frac{\nstep}{\hESS} = \nstep \, \exp\left(k' \, \frac{\ndof}{\nstep}\right),
\end{equation}
using the $k'$ obtained from the fit reported in Fig.~\ref{fig:nemc_6.0_ess_nstep_scaling}. The minimum of the quantity of Eq.~\eqref{eq:eff2} with respect to $\nstep$ leads to an amusing result, i.e., 
\begin{equation}
    \hESS_{\rm best} = \frac{1}{e} = 0.368 \dots,
\end{equation}
which remarkably is architecture-independent (since $k'$ drops out) and also in very good qualitative agreement with what we observe in Fig.~\ref{fig:6.0_eff}.
Hence, in the following we will aim at using flows with $\hESS$ in the 0.2--0.5 range, which appears to be the region where the flows are most efficient.

\section{Sampling topology towards the continuum limit}
\label{sec:continuum}

From the discussion of the previous sections, it is clear that, if we wish to remove the effects introduced by the presence of open boundaries, we can do that with excellent control over the efficiency of the calculation using flow-based approaches such as NE-MCMC and SNFs. Thus, we are finally ready to move to finer lattice spacings and verify whether this family of methodologies can actually be applied to cases where topological slow modes severely affect standard Monte Carlo simulations.

The first step is to include autocorrelations between samples of a given observable in the cost function of Eq.~\eqref{eq:eff}. This is implemented with the integrated autocorrelation time $\tauint(\mathcal{O})$, which naturally leads to a more appropriate definition of the cost effectiveness as
\begin{align}
\label{eq:eff3}
    \Eff_{\mathrm{f}} ( \mathcal{O} ) &\equiv \Var_{\mathrm{f}} ( \mathcal{O} ) \times 2 \tauint(\mathcal{O}) \times (\nstep + \nbetween) \\ &\simeq \Var ( \mathcal{O} )\frac{2 \tauint(\mathcal{O})}{\hESS} (\nstep + \nbetween).
\end{align}

This quantity provides intuition for a possible strategy to efficiently compute topological observables in the continuum limit: namely, we can fix the value of $\ESS$ and $\tauint$ to some desired value by tuning $\nstep$ and $\nbetween$ in a suitable way.  
On the one hand, autocorrelations for $\QL^2$ are expected to scale with $a^{-2}$ in the presence of OBC~\cite{Luscher:2011kk,McGlynn:2014bxa}: hence, by increasing $\nbetween$ in the same fashion we expect to keep the value of $\tauint$ roughly fixed. Of course, we expect this to hold (at least approximately) only if the size of defect $L_\dd$ is kept fixed in physical units as well. As a consequence of this, in the continuum limit $L_\dd/a$ grows and the $\hESS$ is expected to decrease exponentially at fixed $\nstep$.
On the other hand, from the detailed discussion of Section~\ref{sec:scaling}, we have very good control of the relationship between the $\hESS$, $L_d/a$ and $\nstep$. More specifically, by increasing $\nstep$ proportionally to $(L_\dd/a)^3$, the $\ESS$ will be kept (in excellent approximation) fixed\footnote{This assumes that the $\hESS$ is a good approximation of the true $\ESS$ for $\QL^2$, see Eq.~\eqref{eq:ess_true}.}.

The strategy is then fully outlined: one has to scale $\nbetween$ with $a^{-2}$ to keep autocorrelations roughly fixed and $\nstep$ with $a^{-3}$ to maintain the efficiency of the flow intact in the continuum limit. The effective costs will then grow like
\begin{equation}
\label{eq:continuum_scaling}
    \Eff_{\mathrm{f}} (\QL^2) \sim \Var ( \QL^2 ) (k_0 a^{-3}  + k_1 a^{-2}),
\end{equation}
where the coefficients $k_0$ and $k_1$ depend on the specific setup of the flow.

Before the discussion of numerical results at finer lattice spacings, let us take a closer look at this expected scaling in the continuum, in particular at the coefficients we introduced. For instance, $k_1$ will be smaller for defects that are larger in physical units; furthermore, the same coefficient still contains a residual dependence on $\nstep$, as the decorrelation does not occur simply in the prior distribution, but also during the non-equilibrium evolution. Looking at $k_0$ instead, it is clear that it depends heavily on the architecture itself: for example, one can imagine a more efficient coupling layer that further reduces the cost in units of $\nstep$ to reach the same value of $\hESS$. 

In the flows studied in this work, we always have $\nstep > \nbetween$, at least by a factor 3 if not more. It makes sense, then, to ``space'' the evolutions by increasing $\nbetween$, as the largest contribution to the sampling costs comes from the flow itself. However, with the development of more advanced SNF architectures the corresponding coefficient $k_0$ will become much smaller, and the $a^{-2}$ term will be the dominant one. In a sense, in this situation the whole simulation would look more similar to a standard one, with the addition of a lightweight flow that safely removes all OBC effects; the role of $\nbetween$ would also be less relevant, as the flow itself would be cheap to apply. We reckon this is the explicit goal of future developments for SNF-based approaches.

We now turn to some numerical tests we conducted at relatively fine lattice spacings in order to perform a variety of checks for this approach. First, we wish to verify that the autocorrelations are indeed under control; second, that the scaling of both $\neavg{\Wd}$ and $\hESS$ still holds at larger values of $\beta$; third, that the training strategy for SNFs explained in Section~\ref{sec:snf} is viable also in this regime.  

We report in Table~\ref{tab:fine_spacings_setup} the setup of our simulations and in Table~\ref{tab:fine_spacings_results} the details of the flows we used. We applied both the NE-MCMC and the SNF architectures described in detail in Sections~\ref{sec:nemcmc} and \ref{sec:snf} respectively. Each flow is identified by three main parameters: $L_d$, $\nstep$ and $\nbetween$. As a final goal, we also aim to understand which combination of these parameters is the most efficient to sample topological observables, at least to a good approximation.

\begin{table}[!t]
    \centering
    \begin{tabular}{|c|c|c|c|c|c|c|}
    \hline
        $\beta$ & $r_0/a$ & $a$[fm] & $L/a$ & $L$[fm] & $L_d/a$ & $L_d$[fm]\\
        \hline
        6.4     & 9.74    & 0.0485  & 30    & 1.46    & 3, 4     & 0.15, 0.19\\
        6.5     & 11.09   & 0.0426  & 34    & 1.45    & 4, 5     & 0.17, 0.21\\
    \hline
    \end{tabular}
    \caption{Setup of our simulations at finer lattice spacings, with the corresponding volume and defect size in physical and lattice units. In order to set the scale we used Ref.~\cite{Necco:2001xg}.}
    \label{tab:fine_spacings_setup}
\end{table}

\begin{table}[!t]
    \centering
    \begin{tabular}{|c|c|c|c|c|c|c|c|}
    \hline
     $\beta$ & flow & $L_d/a$ &  $\nstep$ & $\nbetween$ & $\nev$ & $\hESS$ & $\tauint(\QL^2)$ \\
    \hline
     \multirow{6}{*}{6.4} 
      & NE-MCMC & 3 & 250 & 100 & 1050 & 0.16(4) & 1.01(14) \\
      & NE-MCMC & 3 & 250 & 200 & 1000 & 0.13(2) & 0.50(3) \\
      & NE-MCMC & 3 & 400 & 100 & 1000 & 0.29(4) & 0.68(7) \\
      & NE-MCMC & 3 & 600 & 100 & 1020 & 0.44(3) & 0.61(7) \\
      & NE-MCMC & 4 & 590 & 50  &  800 & 0.18(5) & 0.54(6) \\
      & NE-MCMC & 4 & 950 & 50  &  960 & 0.33(3) & 0.50(3) \\
      & SNF     & 3 & 200 & 100 & 1080 & 0.41(2) & 1.5(3) \\
      & SNF     & 3 & 600 & 100 & 1200 & 0.74(1) & 0.71(9) \\
    \hline
     \multirow{5}{*}{6.5} 
      & NE-MCMC & 4 & 595  & 130 & 1340 & 0.04(3) & 0.74(7) \\
      & NE-MCMC & 4 & 950  & 130 & 900  & 0.39(4) & 0.75(9) \\
      & NE-MCMC & 4 & 1425 & 130 & 600  & 0.52(2) & 0.67(9) \\
      & NE-MCMC & 5 & 1153 & 65  & 720  & 0.20(2) & 0.5(1) \\
      & NE-MCMC & 5 & 1860 & 65  & 450  & 0.45(2) & 0.5(1) \\
      & SNF     & 4 & 475  & 130 & 1020 & 0.37(3) & 0.9(1) \\
      & SNF     & 4 & 1425 & 130 & 770  & 0.71(2) & 0.5(1) \\
    \hline
    \end{tabular}
    \caption{Details of the various flow architectures used in the simulations at the two finer lattice spacings and the corresponding values of $\hESS$ and $\tauint(\QL^2)$.}
    \label{tab:fine_spacings_results}
\end{table}

\begin{figure}[!t]
    \centering
    \includegraphics[width=\textwidth]{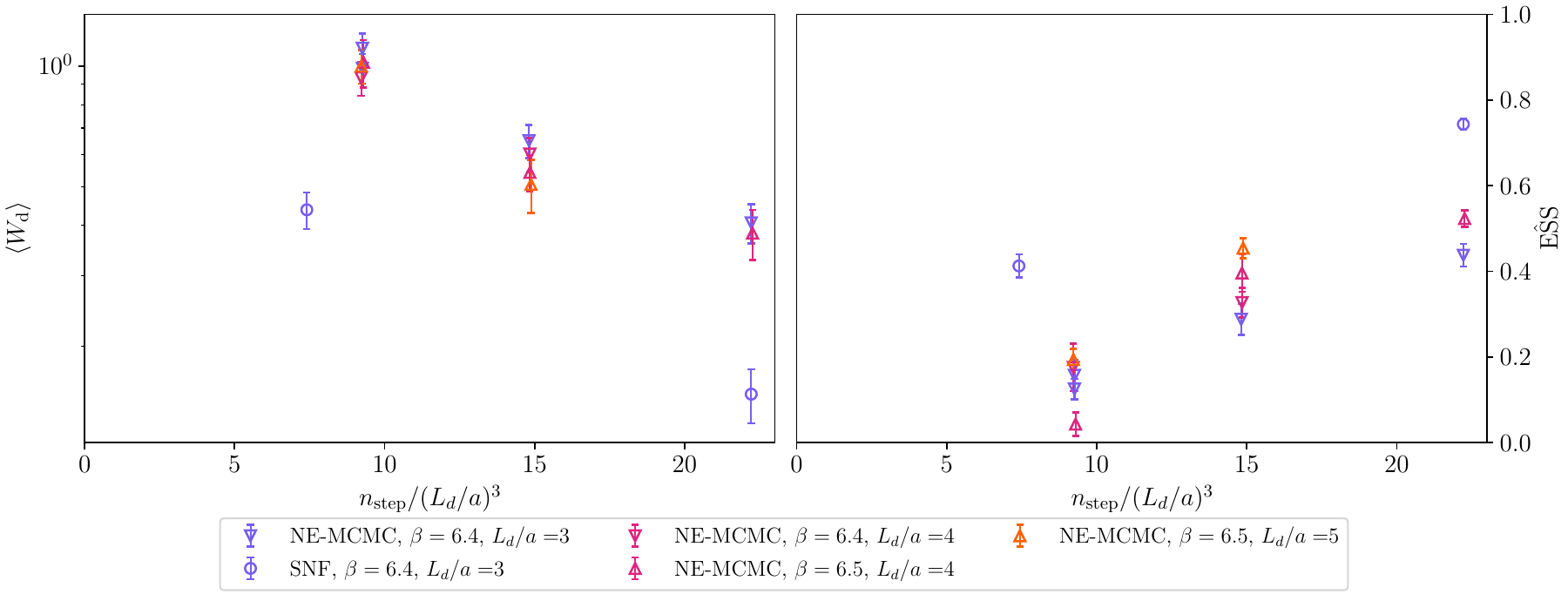}
    \caption{Results for the dissipated work of Eq.~\eqref{eq:kl} (left panel) and the effective sample size of Eq.~\eqref{eq:ess} (right panel) for flows in the boundary conditions as a function of $\nstep$ divided by the size of the defect. All results were obtained either at $\beta=6.4$ on a $30^4$ lattice or at $\beta=6.5$ on a $34^4$ lattice; see Table~\ref{tab:fine_spacings_setup} for more details. Results for $\beta=6.4$ and $L_d/a=3$ obtained with NE-MCMC are connected with a line to help distinguish the scaling of the purely stochastic flow.}
    \label{fig:6.4_6.5_metrics}
\end{figure}

Looking at Table~\ref{tab:fine_spacings_results} we can immediately observe that the autocorrelations of $\QL^2$ are completely under control, with values of $\tauint(\QL^2)$ never significantly larger than 1; this is an unquestionable signal that the topological charge is sampled efficiently for these choices of $L_d$, $\nstep$ and $\nbetween$. Results for $\neavg{W_d}$ and $\hESS$ are also reported in Fig.~\ref{fig:6.4_6.5_metrics}, where it is possible to appreciate the same scaling of $\nstep$ with the number of degrees of freedom varied in the evolution, i.e., $(L_d/a)^3$, that we discussed in Section~\ref{sec:scaling}. Note that the $\ESS$ is a relatively noise metric, since it is the combination of two exponential averages; hence, with the limited statistics obtained on these larger volumes, the collapse over a single curve is less striking than in the $\beta=6.0$ results in Figs.~\ref{fig:6.0_dkl_nemc_snf} and \ref{fig:6.0_ess_nemc_snf}.

We also implemented SNF architectures using the same procedure followed in Section~\ref{sec:snf} for $\beta=6.0$. In particular, we trained the coupling layers only for architectures with $\nstep=8,16$ on a small lattice with
$L/a=16$: the weights were then interpolated using the procedure described in Appendix~\ref{sec:app_rho_interpolation} and applied to the sampling for all architectures on the lattice setups reported in Table~\ref{tab:fine_spacings_setup}. 
While there is no guarantee that this procedure yields the most efficient flow, it appears to provide nonetheless a remarkable improvement over the standard NE-MCMC both for $\beta=6.4$ with a $L_d/a=3$ defect and for $\beta=6.5$ with a $L_d/a=4$ defect.
In one case, we compared SNFs with NE-MCMC fixing $\nstep$, with the former significantly outperforming the latter both in terms of $\neavg{W_d}$ and $\hESS$. Conversely, we also verified the same improvement factor observed in Section~\ref{sec:snf}: namely, the NE-MCMC metrics could be matched with those from SNFs using only \textit{one third} of Monte Carlo steps.
Once more we stress that this improvement factor is obtained for negligible additional sampling costs and requires a cheap and relatively straightforward training of the stout smearing weights.

At this point it is natural to compare our method with a standard MCMC simulation using the cost function of Eq.~\eqref{eq:eff3} and simply setting $\hESS=1$ and $\nstep + \nbetween=1$, since the only relevant variable is the integrated autocorrelation time. A careful study of the autocorrelations of $Q^2$ is provided in Ref.~\cite{Eichhorn:2023uge}, where using the same MCMC update (1HB+4OR) one obtains $\tauint(Q^2) \sim a^{-z}$ with $z \simeq 5.5$. This is in stark comparison with the scaling behaviour of Eq.~\eqref{eq:continuum_scaling} for the algorithm presented in this work (which is, in this specific implementation, still based on the 1HB+4OR update). Both NE-MCMC and SNF are thus expected to greatly improve on standard simulations for fine enough lattices; this raises a practical question, i.e., at which lattice spacing a given flow-based approach becomes advantageous. 
This can be estimated using the appropriate scaling for the cost function $\Eff (Q^2)$ starting from a known value at a coarser spacing, e.g. $\beta \sim 6.4$. For a standard simulation Ref.~\cite{Eichhorn:2023uge} finds $\tauint(Q^2) \sim 200$; for the flow-based approach we consider one of the SNF architectures we investigated ($\nstep=600$, see Table~\ref{tab:fine_spacings_results}). Extrapolating both approaches to $a\to 0$, one obtains that this particular SNF becomes more efficient below $a \sim 0.03$fm: however, we stress that this target can be met at coarser lattices by developing a better SNF architecture, which has the direct effect of decreasing the coefficient of $k_0$ in Eq.~\eqref{eq:continuum_scaling}.

\begin{figure}[!t]
    \centering
    \includegraphics[width=\textwidth]{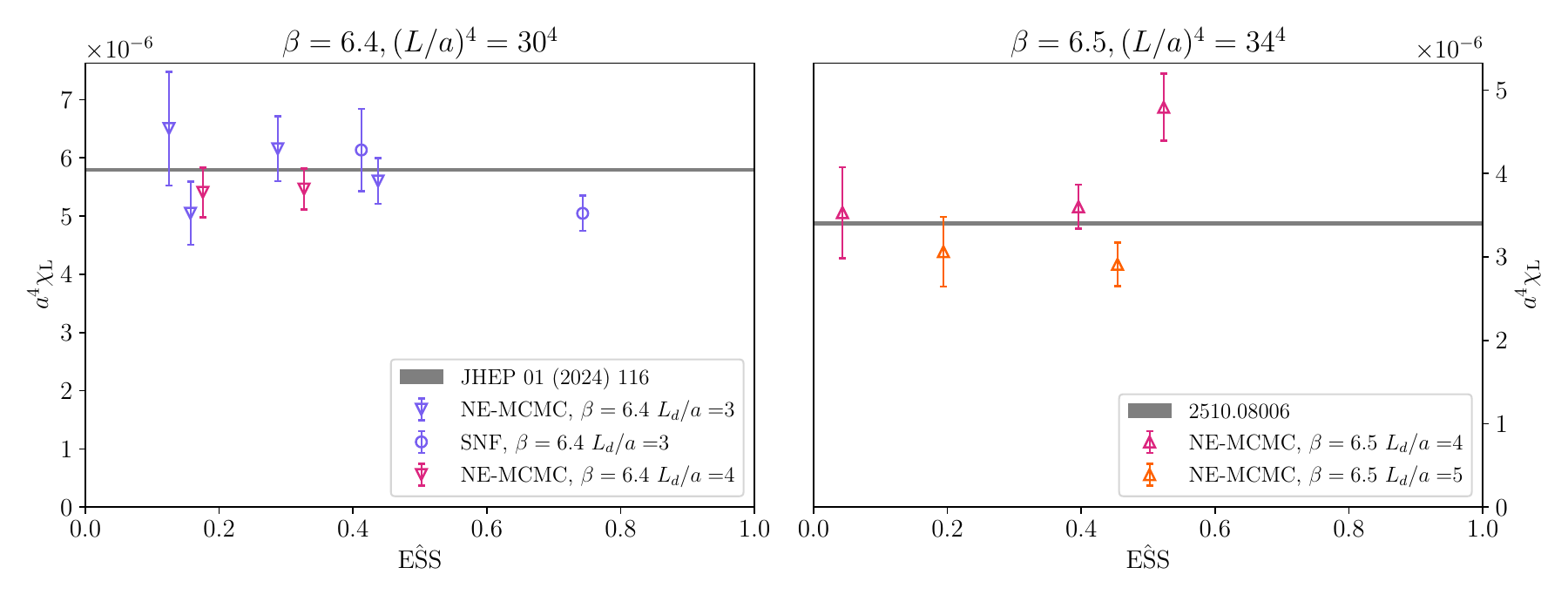}
    \caption{Results for $a^4\chiL$ obtained with the NE-MCMC or SNF architectures described in the text, with $n_{\cool}=60$ ($R_s \simeq 12.6a\sim 0.4 L$). Horizontal bands correspond to results obtained in Refs.~\cite{Bonanno:2023ple,Bonanno:2025eeb}.}
    \label{fig:6.4_6.5_chi}
\end{figure}

Finally, we can look at results for the topological susceptibility: using both NE-MCMC and SNFs, we compute the expectation value of $\QL^2$ appearing in Eq.~\eqref{eq:chi} with the estimator of Eq.~\eqref{eq:estimator} using the appropriate definition of work. In this way we obtain the results shown in Fig.~\ref{fig:6.4_6.5_chi} both for $\beta=6.4$ and $\beta=6.5$, which immediately show perfect agreement with results quoted in Refs.~\cite{Bonanno:2023ple,Bonanno:2025eeb} obtained with much larger statistics. This serves as a sanity check that the methods described in this study do not introduce hidden systematic effects.

\section{Conclusions}\label{sec:conclusions}

In this manuscript we have outlined a flow-based strategy to mitigate topological freezing in lattice gauge theories, based essentially on two ingredients. The first is the use of Open Boundary Conditions, a common tool in lattice calculations that is able to greatly reduce autocorrelations in topological observables. The second, novel ingredient is the use of approaches based on out-of-equilibrium simulations and Normalizing Flows to safely and efficiently remove the unwanted finite-size effects that OBC induce. The combination of the two provides a tool whose scaling in the continuum limit can be well estimated, even if just approximately: extended testing in $\SU(3)$ pure gauge theory both at coarse and fine lattice spacings supports these findings.

We devoted a substantial part of the manuscript to determining the computational cost of applying flows based on out-of-equilibrium evolutions. Indeed, at this stage, this cost still dominates the overall simulation budget towards the continuum, as it scales with $a^{-3}$ for fixed physical defect size; this is to be contrasted with the traditional $a^{-2}$ scaling of the autocorrelations themselves. However, future developments might change the \textit{coefficient} $k_0$ of the $a^{-3}$ scaling in Eq.~\eqref{eq:continuum_scaling}. 
In this work we already worked exactly in this direction, implementing a relatively simple SNF architecture which proved to be a factor 3 more efficient than standard NE-MCMC. We envision that pursuing a systematic improvement program of SNFs, in particular when designing more efficient gauge-equivariant coupling layers, will further decrease the $k_0$ coefficient for a limited cost in terms of training. We plan to do so by building on recent work in this direction~\cite{Abbott:2023thq,Abbott:2024kfc}. 
Acting on the stochastic updates themselves represents a promising direction as well: recent advancements in the literature~\cite{Schmiedl_2007, Gomez_Marin_2008, Sivak_2012, Zulkowski_2012, Rotskoff_2015, Blaber_2020, Blaber_2021, Bonanca_2018, Kamizaki_2022, engel2023optimal} (see Ref.~\cite{Blaber_2023} for a review) provide clear recipes to find an \textit{optimal} protocol given a starting and a target probability distribution, minimizing the dissipated work $\neavg{\Wd}$ given a fixed budget of MCMC steps. We plan to implement these techniques for out-of-equilibrium evolutions in lattice gauge theory: furthermore, achieving this goal would not simply improve efficiency, but enable full control over the fine details of the behavior of this family of flow-based approaches. Fully realizing these improvement programs will simplify the budget of the computation costs: indeed, in the limit of very efficient flows, the simulation would very much look like a standard one, with mild autocorrelations and the effects of OBC swiftly removed.

A natural extension of this work is to probe finer and finer lattice spacings in $\SU(3)$ Yang-Mills theory with correlations in topological observables completely under control. Part of the motivation is theoretical: are the cutoff effects of, e.g., the topological susceptibility under control? Lattice spacings below $0.04$ fm have never been explored for such calculations, as the continuum limit usually relied on the use of measurements obtained on coarser lattices. Furthermore, it is extremely important to understand whether the strategy outlined in this manuscript actually holds in conditions where the standard MCMC features extreme autocorrelations: a large scale simulation of this kind will provide a challenging test.

Interestingly, an extension of this approach to QCD with dynamical fermions presents no particular intrinsic conceptual challenges and would require minor changes to pre-existing codes. The generalization of NE-MCMC is relatively straightforward, requiring the switch to Hybrid Monte Carlo update algorithms instead of the heatbath+overrelaxation combination used in pure gauge simulations. A minimal implementation would also follow the work Ref.~\cite{Bonanno:2024zyn} in PTBC; open boundaries would be set only on the gauge fields, leaving standard antiperiodic ones for fermion fields. Furthermore, most ingredients needed for the design of suitable SNFs have been already studied. Any coupling layer developed for gauge fields in Yang-Mills theory can be directly ported to flow architectures for full QCD; similar transformations for fermionic variables have been also recently developed~\cite{Albergo:2021bna,Abbott:2022zhs}. We leave the study of an optimally performing architecture in the presence of dynamical quark fields to future work.

Finally, we stress once more how the use of flow-based approaches such as NE-MCMC and SNFs can be extended to a broad variety of theoretical setups, well beyond the issue of topological freezing. Recent efforts (including this one) have been focused on systems in which a localized set of degrees of freedom is changed along the evolution, see for example Refs.~\cite{Bulgarelli:2023ofi, Bonanno:2024udh, Bulgarelli:2024yrz, Bulgarelli:2025riv}; in such cases, generally speaking, the only probability distribution of interest is the target one. However, out-of-equilibrium evolutions (and their SNF generalizations) can be naturally applied to setups in which the action of the theory depends on a set of parameters (e.g., quark masses), all of which can be suitably varied (without breaking translational invariance). This is instead a \textit{multicanonical} approach, in which multiple intermediate probability distributions are sampled in the same simulation. A typical example is the computation of the equation of state, in which multiple temperatures are explored within the same evolution, see Ref.~\cite{Caselle:2018kap}. Thus, NE-MCMC and SNFs provide a solid and well-understood framework for a completely different way to perform numerical simulations in lattice gauge theories, which we intend to pursue in our future work.

\acknowledgments
We thank M.~Caselle, G.~Kanwar and M.~Panero for insightful and helpful discussions. The work of C.~B.~is supported by the Spanish Research Agency (Agencia Estatal de Investigación) through the grant IFT Centro de Excelencia Severo Ochoa CEX2020-001007-S and, partially, by grant PID2021-127526NB-I00, both funded by MCIN/AEI/10.13039/ 501100011033. E.~C., A.~N., D.~P. and L.~V. acknowledge support and A.~B. acknowledges partial support by the Simons Foundation grant 994300 (Simons Collaboration on Confinement and QCD Strings).  A.~N. acknowledges support from the European Union - Next Generation EU, Mission 4 Component 1, CUP D53D23002970006, under the Italian PRIN “Progetti di Ricerca di Rilevante Interesse Nazionale – Bando 2022” prot. 2022ZTPK4E. A.~B., E.~C., A.~N., D.~P. and L.~V. acknowledge support from the SFT Scientific Initiative of INFN. The work of D.~V.~is supported by STFC under Consolidated Grant No.~ST/X000680/1. 
We acknowledge EuroHPC Joint Undertaking for awarding the project ID EHPC-DEV-2024D11-010 access to the LEONARDO Supercomputer hosted by the Consorzio Interuniversitario per il Calcolo Automatico dell'Italia Nord Orientale (CINECA), Italy. This work was partially carried out using the computational facilities of the "Lovelace" High Performance Computing Centre, University of Plymouth, https://www.plymouth.ac.uk/about-us/university-structure/faculties/science-engineering/hpc.

\appendix

\section*{Appendix}

\section{Interpolation strategy for defect coupling layer parameters}
\label{sec:app_rho_interpolation}

Every SNF architecture used in this work, after fixing the value of the inverse coupling $\beta$ and the defect size $L_d/a$, should in principle be trained separately for each different value of $\nstep$. However, the protocol we use is the same for all values of $\nstep$, i.e., it is linear in the $\lambda(n)$ parameter appearing in Eq.~\eqref{eq:bc_par}. This means that the path in the intermediate probability distributions is in good approximation the same, with each architecture traversing it at different speeds. Thus, it is not unreasonable to think that the parameters of the defect coupling layers discussed in Section~\ref{sec:snf} belonging to flows with varying $\nstep$ are related to each other; furthermore, the coupling layers used in this work are rather simple, as we are training directly the stout smearing parameters appearing in Eq.~\eqref{eq:cl_staples}. 

Indeed, previous work in Ref.~\cite{Bulgarelli:2024brv} showed that in the case of flow transformations in $\beta$, the stout smearing parameters obtained at a fixed $\nstep^{\rm(train)}$ from the training procedure could be easily interpolated in the index of the coupling layer $l$, with $l \in [1, \nstep^{\rm (train)}]$. The same interpolating function was then used to determine the parameters at much larger values $\nstep$, working remarkably well even for much slower transformations. 

In this work, we perform the same operation, but with a fundamental difference: since the translational and rotational symmetries of the lattice are lost due to the presence of the defect, we train the non-vanishing values of the $\rho^{\pm}_{\mu \nu}(n,x)$ parameters independently. Using separate interpolations would then be impractical, as the number of parameters is already quite large for $L_d/a=2$, see Table~\ref{tab:nrho}. However, the parameters are not really independent, as the links on and around the defect that are interested by the stout smearing transformations still enjoy a residual cubic symmetry. While rigorously implementing the latter would also leave us with a sizable number of parameters, we use a stronger prescription. In particular we consider only 9 ``classes'' of the $\rho^{\pm}_{\mu \nu}(n,x)$ parameters, that we classify with the criteria described in Table~\ref{tab:rho_classification}. 

\begin{table}[t!]
    \centering
    \begin{tabular}{|c|c|c|c|}
    \hline
        class & $U_\mu (x)$ on defect & staple contains defect link & \# other staples containing defect link \\
    \hline
        t1b & Yes & No  & 5 \\
        t1e & Yes & No  & 4 \\
        t1c & Yes & No  & 3 \\
        t2  & Yes & Yes & 5 \\
        t2b & Yes & Yes & 4 \\
        t2e & Yes & Yes & 3 \\
        t2c & Yes & Yes & 2 \\
        sp  & No ($\mu\neq \hat{0}$) & Yes & - \\
        ex  & No ($\mu=\hat{0}$)  & Yes & - \\
    \hline
    \end{tabular}

    \caption{Classification of the stout smearing parameters $\rho^{\pm}_{\mu \nu}(n,x)$ for fixed $n$ obtained from the training described in the main text. $U_\mu(x)$ is the link being transformed in Eq.~\eqref{eq:cl_stout_smearing}, which can be or not be on the defect, see Eq.~\eqref{eq:bc_par}. Each $\rho^{\pm}_{\mu \nu}(n,x)$ multiplies a staple, see Eq.~\ref{eq:cl_staples}, which itself may contain a link on the defect or not. Finally, all the other staples connected to $U_\mu(x)$ may contain a link on the defect as well. If no criteria is met, $\rho^{\pm}_{\mu \nu}(n,x)$ is set to zero.}
    \label{tab:rho_classification}
\end{table}

\begin{figure}[t!]
    \centering
    \includegraphics[width=0.75\linewidth]{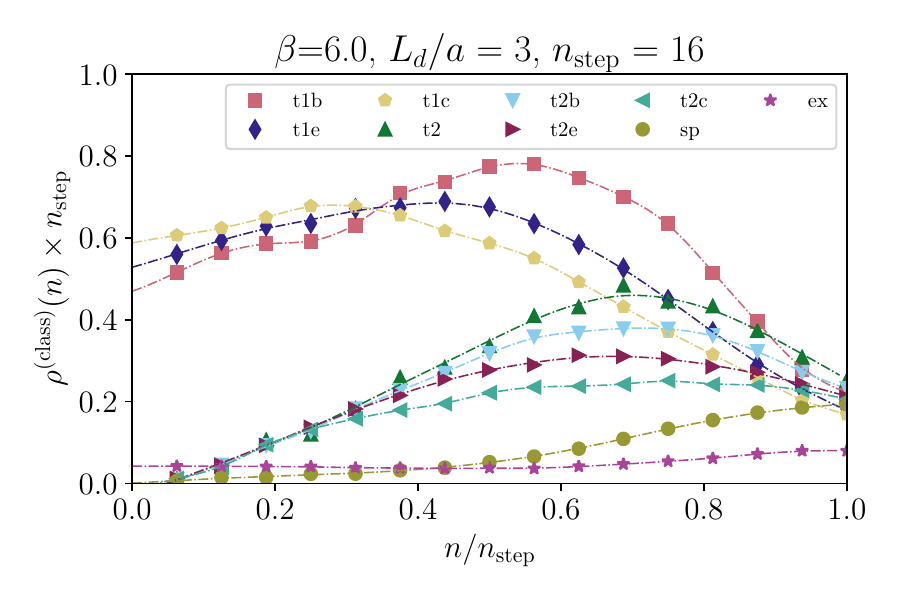}
    \caption{Value of the averaged stout smearing parameters $\rho^{\pm}_{\mu \nu}(n,x)$ classified according to the prescription of Table~\ref{tab:rho_classification} and multiplied by $\nstep$, along the various coupling layers with label $n$.}
    \label{fig:rho_splines}
\end{figure}

This classification was suggested by direct inspection of the relevant smearing parameters in each of these families; we stress that it does not need to be exact, but only be able to transfer the relevant pieces of information obtained from a training at small $\nstep$ to an architecture at larger $\nstep$.
After dividing all the $\rho^{\pm}_{\mu \nu}(n,x)$ values in each of the 9 sets of Table~\ref{tab:rho_classification}, we take the average of all the parameters belonging to that specific class (denoted with $\rho^{\rm (class)}(n)$) and we multiply it by $\nstep$. We show these values in Fig.~\ref{fig:rho_splines} for the case of $\beta=6.0$, $L_d/a=3$ and $\nstep=16$. The behavior for different inverse couplings and defect sizes is qualitatively similar, with the exception of $L_d/a=2$, where only 5 classes can be identified due to geometry constraints. 
The last step is a spline interpolation in $n/\nstep \in (0,1]$ of $\rho^{\rm (class)}(n) \times \nstep$: the resulting function (simply divided by $\nstep$) provides an extrapolation of $\rho^{\rm (class)}(n)$ for any number of steps in the evolution; the corresponding interpolations are showed in Fig.~\ref{fig:rho_splines}.

\bibliographystyle{JHEP}
\bibliography{biblio}

\providecommand{\href}[2]{#2}\begingroup\raggedright\begin{thebibliography}{100}

\bibitem{Alles:1996vn}
B.~Alles, G.~Boyd, M.~D'Elia, A.~Di~Giacomo and E.~Vicari, \emph{{Hybrid Monte
  Carlo and topological modes of full QCD}},
  \href{https://doi.org/10.1016/S0370-2693(96)01247-6}{\emph{Phys. Lett. B}
  {\bfseries 389} (1996) 107}
  [\href{https://arxiv.org/abs/hep-lat/9607049}{{\ttfamily hep-lat/9607049}}].

\bibitem{DelDebbio:2002xa}
L.~Del~Debbio, H.~Panagopoulos and E.~Vicari, \emph{{$\theta$ dependence of
  $SU(N)$ gauge theories}},
  \href{https://doi.org/10.1088/1126-6708/2002/08/044}{\emph{JHEP} {\bfseries
  08} (2002) 044} [\href{https://arxiv.org/abs/hep-th/0204125}{{\ttfamily
  hep-th/0204125}}].

\bibitem{DelDebbio:2004xh}
L.~Del~Debbio, G.~M. Manca and E.~Vicari, \emph{{Critical slowing down of
  topological modes}},
  \href{https://doi.org/10.1016/j.physletb.2004.05.038}{\emph{Phys. Lett. B}
  {\bfseries 594} (2004) 315}
  [\href{https://arxiv.org/abs/hep-lat/0403001}{{\ttfamily hep-lat/0403001}}].

\bibitem{Schaefer:2010hu}
{\scshape ALPHA} collaboration, S.~Schaefer, R.~Sommer and F.~Virotta,
  \emph{{Critical slowing down and error analysis in lattice QCD simulations}},
  \href{https://doi.org/10.1016/j.nuclphysb.2010.11.020}{\emph{Nucl. Phys. B}
  {\bfseries 845} (2011) 93} [\href{https://arxiv.org/abs/1009.5228}{{\ttfamily
  1009.5228}}].

\bibitem{Luscher:1981zq}
M.~L{\"u}scher, \emph{{Topology of Lattice Gauge Fields}},
  \href{https://doi.org/10.1007/BF02029132}{\emph{Commun. Math. Phys.}
  {\bfseries 85} (1982) 39}.

\bibitem{Durr:2006ky}
S.~D{\"u}rr, Z.~Fodor, C.~Hoelbling and T.~Kurth, \emph{{Precision study of the
  SU(3) topological susceptibility in the continuum}},
  \href{https://doi.org/10.1088/1126-6708/2007/04/055}{\emph{JHEP} {\bfseries
  04} (2007) 055} [\href{https://arxiv.org/abs/hep-lat/0612021}{{\ttfamily
  hep-lat/0612021}}].

\bibitem{Ce:2015qha}
M.~C\`e, C.~Consonni, G.~P. Engel and L.~Giusti, \emph{{Non-Gaussianities in
  the topological charge distribution of the $SU(3)$ Yang--Mills theory}},
  \href{https://doi.org/10.1103/PhysRevD.92.074502}{\emph{Phys. Rev. D}
  {\bfseries 92} (2015) 074502}
  [\href{https://arxiv.org/abs/1506.06052}{{\ttfamily 1506.06052}}].

\bibitem{Bonati:2015sqt}
C.~Bonati, M.~D'Elia and A.~Scapellato, \emph{{$\theta$ dependence in $SU(3)$
  Yang-Mills theory from analytic continuation}},
  \href{https://doi.org/10.1103/PhysRevD.93.025028}{\emph{Phys. Rev. D}
  {\bfseries 93} (2016) 025028}
  [\href{https://arxiv.org/abs/1512.01544}{{\ttfamily 1512.01544}}].

\bibitem{Athenodorou:2020ani}
A.~Athenodorou and M.~Teper, \emph{{The glueball spectrum of SU(3) gauge theory
  in 3 + 1 dimensions}},
  \href{https://doi.org/10.1007/JHEP11(2020)172}{\emph{JHEP} {\bfseries 11}
  (2020) 172} [\href{https://arxiv.org/abs/2007.06422}{{\ttfamily
  2007.06422}}].

\bibitem{Bonanno:2023ple}
C.~Bonanno, \emph{{The topological susceptibility slope $\chi^\prime$ of the
  pure-gauge SU(3) Yang-Mills theory}},
  \href{https://doi.org/10.1007/JHEP01(2024)116}{\emph{JHEP} {\bfseries 01}
  (2024) 116} [\href{https://arxiv.org/abs/2311.06646}{{\ttfamily
  2311.06646}}].

\bibitem{Durr:2025qtq}
S.~D{\"u}rr and G.~Fuwa, \emph{{The topological susceptibility and excess
  kurtosis in SU(3) Yang-Mills theory}},
  \href{https://arxiv.org/abs/2501.08217}{{\ttfamily 2501.08217}}.

\bibitem{DelDebbio:2004ns}
L.~Del~Debbio, L.~Giusti and C.~Pica, \emph{{Topological susceptibility in the
  $SU(3)$ gauge theory}},
  \href{https://doi.org/10.1103/PhysRevLett.94.032003}{\emph{Phys. Rev. Lett.}
  {\bfseries 94} (2005) 032003}
  [\href{https://arxiv.org/abs/hep-th/0407052}{{\ttfamily hep-th/0407052}}].

\bibitem{Luscher:2010ik}
M.~L{\"u}scher and F.~Palombi, \emph{{Universality of the topological
  susceptibility in the $SU(3)$ gauge theory}},
  \href{https://doi.org/10.1007/JHEP09(2010)110}{\emph{JHEP} {\bfseries 09}
  (2010) 110} [\href{https://arxiv.org/abs/1008.0732}{{\ttfamily 1008.0732}}].

\bibitem{Cichy:2015jra}
{\scshape ETM} collaboration, K.~Cichy, E.~Garcia-Ramos, K.~Jansen, K.~Ottnad
  and C.~Urbach, \emph{{Non-perturbative Test of the Witten-Veneziano Formula
  from Lattice QCD}},
  \href{https://doi.org/10.1007/JHEP09(2015)020}{\emph{JHEP} {\bfseries 09}
  (2015) 020} [\href{https://arxiv.org/abs/1504.07954}{{\ttfamily
  1504.07954}}].

\bibitem{Bonanno:2019xhg}
C.~Bonanno, G.~Clemente, M.~D'Elia and F.~Sanfilippo, \emph{{Topology via
  spectral projectors with staggered fermions}},
  \href{https://doi.org/10.1007/JHEP10(2019)187}{\emph{JHEP} {\bfseries 10}
  (2019) 187} [\href{https://arxiv.org/abs/1908.11832}{{\ttfamily
  1908.11832}}].

\bibitem{Ce:2016awn}
M.~C\`e, M.~Garcia~Vera, L.~Giusti and S.~Schaefer, \emph{{The topological
  susceptibility in the large-$N$ limit of SU($N$) Yang-Mills theory}},
  \href{https://doi.org/10.1016/j.physletb.2016.09.029}{\emph{Phys. Lett. B}
  {\bfseries 762} (2016) 232}
  [\href{https://arxiv.org/abs/1607.05939}{{\ttfamily 1607.05939}}].

\bibitem{Bonati:2016tvi}
C.~Bonati, M.~D'Elia, P.~Rossi and E.~Vicari, \emph{{$\theta$ dependence of 4D
  $SU(N)$ gauge theories in the large-$N$ limit}},
  \href{https://doi.org/10.1103/PhysRevD.94.085017}{\emph{Phys. Rev. D}
  {\bfseries 94} (2016) 085017}
  [\href{https://arxiv.org/abs/1607.06360}{{\ttfamily 1607.06360}}].

\bibitem{Bonanno:2020hht}
C.~Bonanno, C.~Bonati and M.~D'Elia, \emph{{Large-$N$ $SU(N)$ Yang-Mills
  theories with milder topological freezing}},
  \href{https://doi.org/10.1007/JHEP03(2021)111}{\emph{JHEP} {\bfseries 03}
  (2021) 111} [\href{https://arxiv.org/abs/2012.14000}{{\ttfamily
  2012.14000}}].

\bibitem{Athenodorou:2021qvs}
A.~Athenodorou and M.~Teper, \emph{{SU(N) gauge theories in 3+1 dimensions:
  glueball spectrum, string tensions and topology}},
  \href{https://doi.org/10.1007/JHEP12(2021)082}{\emph{JHEP} {\bfseries 12}
  (2021) 082} [\href{https://arxiv.org/abs/2106.00364}{{\ttfamily
  2106.00364}}].

\bibitem{Brower:2003yx}
R.~Brower, S.~Chandrasekharan, J.~W. Negele and U.~J. Wiese, \emph{{QCD at
  fixed topology}},
  \href{https://doi.org/10.1016/S0370-2693(03)00369-1}{\emph{Phys. Lett. B}
  {\bfseries 560} (2003) 64}
  [\href{https://arxiv.org/abs/hep-lat/0302005}{{\ttfamily hep-lat/0302005}}].

\bibitem{Aoki:2007ka}
S.~Aoki, H.~Fukaya, S.~Hashimoto and T.~Onogi, \emph{{Finite volume QCD at
  fixed topological charge}},
  \href{https://doi.org/10.1103/PhysRevD.76.054508}{\emph{Phys. Rev. D}
  {\bfseries 76} (2007) 054508}
  [\href{https://arxiv.org/abs/0707.0396}{{\ttfamily 0707.0396}}].

\bibitem{Fritzsch:2013yxa}
P.~Fritzsch, A.~Ramos and F.~Stollenwerk, \emph{{Critical slowing down and the
  gradient flow coupling in the Schr\"odinger functional}},
  \href{https://doi.org/10.22323/1.187.0461}{\emph{PoS} {\bfseries LATTICE2013}
  (2014) 461} [\href{https://arxiv.org/abs/1311.7304}{{\ttfamily 1311.7304}}].

\bibitem{RamosMartinez:2023tvx}
A.~Ramos~Martinez and G.~Catumba, \emph{{Testing universality of gauge
  theories}}, \href{https://doi.org/10.22323/1.430.0383}{\emph{PoS} {\bfseries
  LATTICE2022} (2023) 383}.

\bibitem{Bietenholz:2015rsa}
W.~Bietenholz, P.~de~Forcrand and U.~Gerber, \emph{{Topological Susceptibility
  from Slabs}}, \href{https://doi.org/10.1007/JHEP12(2015)070}{\emph{JHEP}
  {\bfseries 12} (2015) 070}
  [\href{https://arxiv.org/abs/1509.06433}{{\ttfamily 1509.06433}}].

\bibitem{Laio:2015era}
A.~Laio, G.~Martinelli and F.~Sanfilippo, \emph{{Metadynamics surfing on
  topology barriers: the $CP^{N-1}$ case}},
  \href{https://doi.org/10.1007/JHEP07(2016)089}{\emph{JHEP} {\bfseries 07}
  (2016) 089} [\href{https://arxiv.org/abs/1508.07270}{{\ttfamily
  1508.07270}}].

\bibitem{Luscher:2017cjh}
M.~L\"uscher, \emph{{Stochastic locality and master-field simulations of very
  large lattices}},
  \href{https://doi.org/10.1051/epjconf/201817501002}{\emph{EPJ Web Conf.}
  {\bfseries 175} (2018) 01002}
  [\href{https://arxiv.org/abs/1707.09758}{{\ttfamily 1707.09758}}].

\bibitem{Bonati:2017woi}
C.~Bonati and M.~D'Elia, \emph{{Topological critical slowing down: variations
  on a toy model}},
  \href{https://doi.org/10.1103/PhysRevE.98.013308}{\emph{Phys. Rev. E}
  {\bfseries 98} (2018) 013308}
  [\href{https://arxiv.org/abs/1709.10034}{{\ttfamily 1709.10034}}].

\bibitem{Giusti:2018cmp}
L.~Giusti and M.~L\"uscher, \emph{{Topological susceptibility at $T>T_{\rm c}$
  from master-field simulations of the SU(3) gauge theory}},
  \href{https://doi.org/10.1140/epjc/s10052-019-6706-7}{\emph{Eur. Phys. J. C}
  {\bfseries 79} (2019) 207}
  [\href{https://arxiv.org/abs/1812.02062}{{\ttfamily 1812.02062}}].

\bibitem{Florio:2019nte}
A.~Florio, O.~Kaczmarek and L.~Mazur, \emph{{Open-Boundary Conditions in the
  Deconfined Phase}},
  \href{https://doi.org/10.1140/epjc/s10052-019-7564-z}{\emph{Eur. Phys. J. C}
  {\bfseries 79} (2019) 1039}
  [\href{https://arxiv.org/abs/1903.02894}{{\ttfamily 1903.02894}}].

\bibitem{Funcke:2019zna}
L.~Funcke, K.~Jansen and S.~K\"uhn, \emph{{Topological vacuum structure of the
  Schwinger model with matrix product states}},
  \href{https://doi.org/10.1103/PhysRevD.101.054507}{\emph{Phys. Rev. D}
  {\bfseries 101} (2020) 054507}
  [\href{https://arxiv.org/abs/1908.00551}{{\ttfamily 1908.00551}}].

\bibitem{Albandea:2021lvl}
D.~Albandea, P.~Hern\'andez, A.~Ramos and F.~Romero-L\'opez, \emph{{Topological
  sampling through windings}},
  \href{https://doi.org/10.1140/epjc/s10052-021-09677-6}{\emph{Eur. Phys. J. C}
  {\bfseries 81} (2021) 873}
  [\href{https://arxiv.org/abs/2106.14234}{{\ttfamily 2106.14234}}].

\bibitem{Cossu:2021bgn}
G.~Cossu, D.~Lancastera, B.~Lucini, R.~Pellegrini and A.~Rago, \emph{{Ergodic
  sampling of the topological charge using the density of states}},
  \href{https://doi.org/10.1140/epjc/s10052-021-09161-1}{\emph{Eur. Phys. J. C}
  {\bfseries 81} (2021) 375}
  [\href{https://arxiv.org/abs/2102.03630}{{\ttfamily 2102.03630}}].

\bibitem{Borsanyi:2021gqg}
S.~Borsanyi and D.~Sexty, \emph{{Topological susceptibility of pure gauge
  theory using Density of States}},
  \href{https://doi.org/10.1016/j.physletb.2021.136148}{\emph{Phys. Lett. B}
  {\bfseries 815} (2021) 136148}
  [\href{https://arxiv.org/abs/2101.03383}{{\ttfamily 2101.03383}}].

\bibitem{Fritzsch:2021klm}
P.~Fritzsch, J.~Bulava, M.~C\`e, A.~Francis, M.~L\"uscher and A.~Rago,
  \emph{{Master-field simulations of QCD}},
  \href{https://doi.org/10.22323/1.396.0465}{\emph{PoS} {\bfseries LATTICE2021}
  (2022) 465} [\href{https://arxiv.org/abs/2111.11544}{{\ttfamily
  2111.11544}}].

\bibitem{Eichhorn:2023uge}
T.~Eichhorn, G.~Fuwa, C.~Hoelbling and L.~Varnhorst, \emph{{Parallel tempered
  metadynamics: Overcoming potential barriers without surfing or tunneling}},
  \href{https://doi.org/10.1103/PhysRevD.109.114504}{\emph{Phys. Rev. D}
  {\bfseries 109} (2024) 114504}
  [\href{https://arxiv.org/abs/2307.04742}{{\ttfamily 2307.04742}}].

\bibitem{Howarth:2023bwk}
D.~Howarth and A.~J. Peterson, \emph{{Topological charge unfreezing with
  AMReX}},  \href{https://arxiv.org/abs/2312.11599}{{\ttfamily 2312.11599}}.

\bibitem{Albandea:2024fui}
D.~Albandea, G.~Catumba and A.~Ramos, \emph{{Strong CP problem in the quantum
  rotor}}, \href{https://doi.org/10.1103/PhysRevD.110.094512}{\emph{Phys. Rev.
  D} {\bfseries 110} (2024) 094512}
  [\href{https://arxiv.org/abs/2402.17518}{{\ttfamily 2402.17518}}].

\bibitem{Abe:2024fpt}
M.~Abe, O.~Morikawa and H.~Suzuki, \emph{{Monte Carlo Simulation of the
  $\mathrm{SU}(2)/\mathbb{Z}_2$ Yang--Mills Theory}},
  \href{https://doi.org/10.1093/ptep/ptaf075}{\emph{PTEP} {\bfseries 2025}
  (2025) 6, 063B03} [\href{https://arxiv.org/abs/2501.00286}{{\ttfamily
  2501.00286}}].

\bibitem{Finkenrath:2023sjg}
J.~Finkenrath, \emph{{Review on Algorithms for dynamical fermions}},
  \href{https://doi.org/10.22323/1.430.0227}{\emph{PoS} {\bfseries LATTICE2022}
  (2023) 227} [\href{https://arxiv.org/abs/2402.11704}{{\ttfamily
  2402.11704}}].

\bibitem{Boyle:2024nlh}
P.~A. Boyle, \emph{{Advances in algorithms for solvers and gauge generation}},
  \href{https://doi.org/10.22323/1.453.0122}{\emph{PoS} {\bfseries LATTICE2023}
  (2024) 122} [\href{https://arxiv.org/abs/2401.16620}{{\ttfamily
  2401.16620}}].

\bibitem{Finkenrath:2024ptc}
J.~Finkenrath, \emph{{Future trends in lattice QCD simulations}},
  \href{https://doi.org/10.22323/1.451.0009}{\emph{PoS} {\bfseries
  EuroPLEx2023} (2024) 009}.

\bibitem{Luscher:2011kk}
M.~L{\"u}scher and S.~Schaefer, \emph{{Lattice QCD without topology barriers}},
  \href{https://doi.org/10.1007/JHEP07(2011)036}{\emph{JHEP} {\bfseries 07}
  (2011) 036} [\href{https://arxiv.org/abs/1105.4749}{{\ttfamily 1105.4749}}].

\bibitem{Luscher:2012av}
M.~Luscher and S.~Schaefer, \emph{{Lattice QCD with open boundary conditions
  and twisted-mass reweighting}},
  \href{https://doi.org/10.1016/j.cpc.2012.10.003}{\emph{Comput. Phys. Commun.}
  {\bfseries 184} (2013) 519}
  [\href{https://arxiv.org/abs/1206.2809}{{\ttfamily 1206.2809}}].

\bibitem{McGlynn:2014bxa}
G.~McGlynn and R.~D. Mawhinney, \emph{{Diffusion of topological charge in
  lattice QCD simulations}},
  \href{https://doi.org/10.1103/PhysRevD.90.074502}{\emph{Phys. Rev. D}
  {\bfseries 90} (2014) 074502}
  [\href{https://arxiv.org/abs/1406.4551}{{\ttfamily 1406.4551}}].

\bibitem{Hasenbusch:2017unr}
M.~Hasenbusch, \emph{{Fighting topological freezing in the two-dimensional
  $CP^{N-1}$ model}},
  \href{https://doi.org/10.1103/PhysRevD.96.054504}{\emph{Phys. Rev. D}
  {\bfseries 96} (2017) 054504}
  [\href{https://arxiv.org/abs/1706.04443}{{\ttfamily 1706.04443}}].

\bibitem{Berni:2019bch}
M.~Berni, C.~Bonanno and M.~D'Elia, \emph{{Large-$N$ expansion and
  $\theta$-dependence of $2d$ $CP^{N-1}$ models beyond the leading order}},
  \href{https://doi.org/10.1103/PhysRevD.100.114509}{\emph{Phys. Rev. D}
  {\bfseries 100} (2019) 114509}
  [\href{https://arxiv.org/abs/1911.03384}{{\ttfamily 1911.03384}}].

\bibitem{Bonanno:2022hmz}
C.~Bonanno, \emph{{Lattice determination of the topological susceptibility
  slope $\chi^\prime$ of $2d$ CP$^{N-1}$ models at large $N$}},
  \href{https://doi.org/10.1103/PhysRevD.107.014514}{\emph{Phys. Rev. D}
  {\bfseries 107} (2023) 014514}
  [\href{https://arxiv.org/abs/2212.02330}{{\ttfamily 2212.02330}}].

\bibitem{Bonanno:2022yjr}
C.~Bonanno, M.~D'Elia, B.~Lucini and D.~Vadacchino, \emph{{Towards glueball
  masses of large-N SU(N) pure-gauge theories without topological freezing}},
  \href{https://doi.org/10.1016/j.physletb.2022.137281}{\emph{Phys. Lett. B}
  {\bfseries 833} (2022) 137281}
  [\href{https://arxiv.org/abs/2205.06190}{{\ttfamily 2205.06190}}].

\bibitem{Bonanno:2023hhp}
C.~Bonanno, M.~D'Elia and L.~Verzichelli, \emph{{The $\theta$-dependence of the
  SU(N) critical temperature at large N}},
  \href{https://doi.org/10.1007/JHEP02(2024)156}{\emph{JHEP} {\bfseries 02}
  (2024) 156} [\href{https://arxiv.org/abs/2312.12202}{{\ttfamily
  2312.12202}}].

\bibitem{Bonanno:2024nba}
C.~Bonanno, J.~L. Dasilva~Gol{\'a}n, M.~D'Elia, M.~Garc{\'\i}a~P{\'e}rez and
  A.~Giorgieri, \emph{{The ${\textrm{SU}}(3)$ twisted gradient flow strong
  coupling without topological freezing}},
  \href{https://doi.org/10.1140/epjc/s10052-024-13261-z}{\emph{Eur. Phys. J. C}
  {\bfseries 84} (2024) 916}
  [\href{https://arxiv.org/abs/2403.13607}{{\ttfamily 2403.13607}}].

\bibitem{Bonanno:2024ggk}
C.~Bonanno, C.~Bonati, M.~Papace and D.~Vadacchino, \emph{{The
  $\theta$-dependence of the Yang-Mills spectrum from analytic continuation}},
  \href{https://doi.org/10.1007/JHEP05(2024)163}{\emph{JHEP} {\bfseries 05}
  (2024) 163} [\href{https://arxiv.org/abs/2402.03096}{{\ttfamily
  2402.03096}}].

\bibitem{Bonanno:2024zyn}
C.~Bonanno, G.~Clemente, M.~D'Elia, L.~Maio and L.~Parente, \emph{{Full QCD
  with milder topological freezing}},
  \href{https://doi.org/10.1007/JHEP08(2024)236}{\emph{JHEP} {\bfseries 08}
  (2024) 236} [\href{https://arxiv.org/abs/2404.14151}{{\ttfamily
  2404.14151}}].

\bibitem{Bonanno:2024udh}
C.~Bonanno, A.~Nada and D.~Vadacchino, \emph{{Mitigating topological freezing
  using out-of-equilibrium simulations}},
  \href{https://doi.org/10.1007/JHEP04(2024)126}{\emph{JHEP} {\bfseries 04}
  (2024) 126} [\href{https://arxiv.org/abs/2402.06561}{{\ttfamily
  2402.06561}}].

\bibitem{Bulgarelli:2024brv}
A.~Bulgarelli, E.~Cellini and A.~Nada, \emph{{Scaling of stochastic normalizing
  flows in SU(3) lattice gauge theory}},
  \href{https://doi.org/10.1103/PhysRevD.111.074517}{\emph{Phys. Rev. D}
  {\bfseries 111} (2025) 074517}
  [\href{https://arxiv.org/abs/2412.00200}{{\ttfamily 2412.00200}}].

\bibitem{Caselle:2016wsw}
M.~Caselle, G.~Costagliola, A.~Nada, M.~Panero and A.~Toniato,
  \emph{{Jarzynski\textquoteright{}s theorem for lattice gauge theory}},
  \href{https://doi.org/10.1103/PhysRevD.94.034503}{\emph{Phys. Rev. D}
  {\bfseries 94} (2016) 034503}
  [\href{https://arxiv.org/abs/1604.05544}{{\ttfamily 1604.05544}}].

\bibitem{Caselle:2022acb}
M.~Caselle, E.~Cellini, A.~Nada and M.~Panero, \emph{{Stochastic normalizing
  flows as non-equilibrium transformations}},
  \href{https://doi.org/10.1007/JHEP07(2022)015}{\emph{JHEP} {\bfseries 07}
  (2022) 015} [\href{https://arxiv.org/abs/2201.08862}{{\ttfamily
  2201.08862}}].

\bibitem{Albergo:2019eim}
M.~S. Albergo, G.~Kanwar and P.~E. Shanahan, \emph{{Flow-based generative
  models for Markov chain Monte Carlo in lattice field theory}},
  \href{https://doi.org/10.1103/PhysRevD.100.034515}{\emph{Phys. Rev. D}
  {\bfseries 100} (2019) 034515}
  [\href{https://arxiv.org/abs/1904.12072}{{\ttfamily 1904.12072}}].

\bibitem{Cranmer:2023xbe}
K.~Cranmer, G.~Kanwar, S.~Racani\`ere, D.~J. Rezende and P.~E. Shanahan,
  \emph{{Advances in machine-learning-based sampling motivated by lattice
  quantum chromodynamics}},
  \href{https://doi.org/10.1038/s42254-023-00616-w}{\emph{Nature Rev. Phys.}
  {\bfseries 5} (2023) 526} [\href{https://arxiv.org/abs/2309.01156}{{\ttfamily
  2309.01156}}].

\bibitem{Luscher:2009eq}
M.~L{\"u}scher, \emph{{Trivializing maps, the Wilson flow and the HMC
  algorithm}}, \href{https://doi.org/10.1007/s00220-009-0953-7}{\emph{Commun.
  Math. Phys.} {\bfseries 293} (2010) 899}
  [\href{https://arxiv.org/abs/0907.5491}{{\ttfamily 0907.5491}}].

\bibitem{Engel:2011re}
G.~P. Engel and S.~Schaefer, \emph{{Testing trivializing maps in the Hybrid
  Monte Carlo algorithm}},
  \href{https://doi.org/10.1016/j.cpc.2011.05.004}{\emph{Comput. Phys. Commun.}
  {\bfseries 182} (2011) 2107}
  [\href{https://arxiv.org/abs/1102.1852}{{\ttfamily 1102.1852}}].

\bibitem{rezende2015}
D.~Rezende and S.~Mohamed, \emph{Variational inference with normalizing flows},
   in \emph{Proceedings of the 32nd International Conference on Machine
  Learning}, vol.~37 of \emph{Proceedings of Machine Learning Research},
  (Lille, France), pp.~1530--1538, PMLR, 07--09 Jul, 2015,
  \href{https://arxiv.org/abs/1505.05770}{{\ttfamily 1505.05770}},
  \href{https://proceedings.mlr.press/v37/rezende15.html}{https://proceedings.mlr.press/v37/rezende15.html}.

\bibitem{papamakarios2021}
G.~Papamakarios, E.~T. Nalisnick, D.~J. Rezende, S.~Mohamed and
  B.~Lakshminarayanan, \emph{{Normalizing Flows for Probabilistic Modeling and
  Inference}}, {\emph{J. Mach. Learn. Res.} {\bfseries 22} (2021) 1}.

\bibitem{Nicoli:2020njz}
K.~A. Nicoli, C.~J. Anders, L.~Funcke, T.~Hartung, K.~Jansen, P.~Kessel et~al.,
  \emph{{Estimation of Thermodynamic Observables in Lattice Field Theories with
  Deep Generative Models}},
  \href{https://doi.org/10.1103/PhysRevLett.126.032001}{\emph{Phys. Rev. Lett.}
  {\bfseries 126} (2021) 032001}
  [\href{https://arxiv.org/abs/2007.07115}{{\ttfamily 2007.07115}}].

\bibitem{Nicoli:2019gun}
K.~A. Nicoli, S.~Nakajima, N.~Strodthoff, W.~Samek, K.-R. M{\"u}ller and
  P.~Kessel, \emph{{Asymptotically unbiased estimation of physical observables
  with neural samplers}},
  \href{https://doi.org/10.1103/PhysRevE.101.023304}{\emph{Phys. Rev. E}
  {\bfseries 101} (2020) 023304}
  [\href{https://arxiv.org/abs/1910.13496}{{\ttfamily 1910.13496}}].

\bibitem{DelDebbio:2021qwf}
L.~Del~Debbio, J.~M. Rossney and M.~Wilson, \emph{{Efficient modeling of
  trivializing maps for lattice $\phi^4$ theory using normalizing flows: A
  first look at scalability}},
  \href{https://doi.org/10.1103/PhysRevD.104.094507}{\emph{Phys. Rev. D}
  {\bfseries 104} (2021) 094507}
  [\href{https://arxiv.org/abs/2105.12481}{{\ttfamily 2105.12481}}].

\bibitem{Gerdes:2022eve}
M.~Gerdes, P.~de~Haan, C.~Rainone, R.~Bondesan and M.~C.~N. Cheng,
  \emph{{Learning lattice quantum field theories with equivariant continuous
  flows}}, \href{https://doi.org/10.21468/SciPostPhys.15.6.238}{\emph{SciPost
  Phys.} {\bfseries 15} (2023) 238}
  [\href{https://arxiv.org/abs/2207.00283}{{\ttfamily 2207.00283}}].

\bibitem{Singha:2022icw}
A.~Singha, D.~Chakrabarti and V.~Arora, \emph{{Conditional normalizing flow for
  Markov chain Monte~Carlo sampling in the critical region of lattice field
  theory}}, \href{https://doi.org/10.1103/PhysRevD.107.014512}{\emph{Phys. Rev.
  D} {\bfseries 107} (2023) 014512}
  [\href{https://arxiv.org/abs/2207.00980}{{\ttfamily 2207.00980}}].

\bibitem{Chen:2022ytr}
S.~Chen, O.~Savchuk, S.~Zheng, B.~Chen, H.~Stoecker, L.~Wang et~al.,
  \emph{{Fourier-flow model generating Feynman paths}},
  \href{https://doi.org/10.1103/PhysRevD.107.056001}{\emph{Phys. Rev. D}
  {\bfseries 107} (2023) 056001}
  [\href{https://arxiv.org/abs/2211.03470}{{\ttfamily 2211.03470}}].

\bibitem{Caselle:2023mvh}
M.~Caselle, E.~Cellini and A.~Nada, \emph{{Sampling the lattice Nambu-Goto
  string using Continuous Normalizing Flows}},
  \href{https://doi.org/10.1007/JHEP02(2024)048}{\emph{JHEP} {\bfseries 02}
  (2024) 048} [\href{https://arxiv.org/abs/2307.01107}{{\ttfamily
  2307.01107}}].

\bibitem{Albandea:2023wgd}
D.~Albandea, L.~Del~Debbio, P.~Hern\'andez, R.~Kenway, J.~Marsh~Rossney and
  A.~Ramos, \emph{{Learning trivializing flows}},
  \href{https://doi.org/10.1140/epjc/s10052-023-11838-8}{\emph{Eur. Phys. J. C}
  {\bfseries 83} (2023) 676}
  [\href{https://arxiv.org/abs/2302.08408}{{\ttfamily 2302.08408}}].

\bibitem{Kreit:2025cos}
J.~Kreit, D.~Schuh, K.~A. Nicoli and L.~Funcke, \emph{{SESaMo:
  Symmetry-Enforcing Stochastic Modulation for Normalizing Flows}},
  \href{https://arxiv.org/abs/2505.19619}{{\ttfamily 2505.19619}}.

\bibitem{Schuh:2025gks}
D.~Schuh, J.~Kreit, E.~Berkowitz, L.~Funcke, T.~Luu, K.~A. Nicoli et~al.,
  \emph{{Simulating Correlated Electrons with Symmetry-Enforced Normalizing
  Flows}},  \href{https://arxiv.org/abs/2506.17015}{{\ttfamily 2506.17015}}.

\bibitem{Kanwar:2020xzo}
G.~Kanwar, M.~S. Albergo, D.~Boyda, K.~Cranmer, D.~C. Hackett, S.~Racani\`ere
  et~al., \emph{{Equivariant flow-based sampling for lattice gauge theory}},
  \href{https://doi.org/10.1103/PhysRevLett.125.121601}{\emph{Phys. Rev. Lett.}
  {\bfseries 125} (2020) 121601}
  [\href{https://arxiv.org/abs/2003.06413}{{\ttfamily 2003.06413}}].

\bibitem{Boyda:2020hsi}
D.~Boyda, G.~Kanwar, S.~Racani\`ere, D.~J. Rezende, M.~S. Albergo, K.~Cranmer
  et~al., \emph{{Sampling using $SU(N)$ gauge equivariant flows}},
  \href{https://doi.org/10.1103/PhysRevD.103.074504}{\emph{Phys. Rev. D}
  {\bfseries 103} (2021) 074504}
  [\href{https://arxiv.org/abs/2008.05456}{{\ttfamily 2008.05456}}].

\bibitem{Favoni:2020reg}
M.~Favoni, A.~Ipp, D.~I. M{\"u}ller and D.~Schuh, \emph{{Lattice Gauge
  Equivariant Convolutional Neural Networks}},
  \href{https://doi.org/10.1103/PhysRevLett.128.032003}{\emph{Phys. Rev. Lett.}
  {\bfseries 128} (2022) 032003}
  [\href{https://arxiv.org/abs/2012.12901}{{\ttfamily 2012.12901}}].

\bibitem{Bacchio:2022}
S.~Bacchio, P.~Kessel, S.~Schaefer and L.~Vaitl, \emph{{Learning trivializing
  gradient flows for lattice gauge theories}},
  \href{https://doi.org/10.1103/PhysRevD.107.L051504}{\emph{Phys. Rev. D}
  {\bfseries 107} (2023) L051504}
  [\href{https://arxiv.org/abs/2212.08469}{{\ttfamily 2212.08469}}].

\bibitem{Singha:2023xxq}
A.~Singha, D.~Chakrabarti and V.~Arora, \emph{{Sampling U(1) gauge theory using
  a retrainable conditional flow-based model}},
  \href{https://doi.org/10.1103/PhysRevD.108.074518}{\emph{Phys. Rev. D}
  {\bfseries 108} (2023) 074518}
  [\href{https://arxiv.org/abs/2306.00581}{{\ttfamily 2306.00581}}].

\bibitem{Abbott:2023thq}
R.~Abbott et~al., \emph{{Normalizing flows for lattice gauge theory in
  arbitrary space-time dimension}},
  \href{https://arxiv.org/abs/2305.02402}{{\ttfamily 2305.02402}}.

\bibitem{Gerdes:2024rjk}
M.~Gerdes, P.~de~Haan, R.~Bondesan and M.~C.~N. Cheng, \emph{{Nonperturbative
  trivializing flows for lattice gauge theories}},
  \href{https://doi.org/10.1103/31d5-hvp6}{\emph{Phys. Rev. D} {\bfseries 112}
  (2025) 094516} [\href{https://arxiv.org/abs/2410.13161}{{\ttfamily
  2410.13161}}].

\bibitem{Albergo:2021bna}
M.~S. Albergo, G.~Kanwar, S.~Racani\`ere, D.~J. Rezende, J.~M. Urban, D.~Boyda
  et~al., \emph{{Flow-based sampling for fermionic lattice field theories}},
  \href{https://doi.org/10.1103/PhysRevD.104.114507}{\emph{Phys. Rev. D}
  {\bfseries 104} (2021) 114507}
  [\href{https://arxiv.org/abs/2106.05934}{{\ttfamily 2106.05934}}].

\bibitem{Finkenrath:2022ogg}
J.~Finkenrath, \emph{{Tackling critical slowing down using global correction
  steps with equivariant flows: the case of the Schwinger model}},
  \href{https://arxiv.org/abs/2201.02216}{{\ttfamily 2201.02216}}.

\bibitem{Albergo:2022qfi}
M.~S. Albergo, D.~Boyda, K.~Cranmer, D.~C. Hackett, G.~Kanwar, S.~Racani\`ere
  et~al., \emph{{Flow-based sampling in the lattice Schwinger model at
  criticality}}, \href{https://doi.org/10.1103/PhysRevD.106.014514}{\emph{Phys.
  Rev. D} {\bfseries 106} (2022) 014514}
  [\href{https://arxiv.org/abs/2202.11712}{{\ttfamily 2202.11712}}].

\bibitem{Abbott:2022zhs}
R.~Abbott et~al., \emph{{Gauge-equivariant flow models for sampling in lattice
  field theories with pseudofermions}},
  \href{https://doi.org/10.1103/PhysRevD.106.074506}{\emph{Phys. Rev. D}
  {\bfseries 106} (2022) 074506}
  [\href{https://arxiv.org/abs/2207.08945}{{\ttfamily 2207.08945}}].

\bibitem{Abbott:2024kfc}
R.~Abbott, A.~Botev, D.~Boyda, D.~C. Hackett, G.~Kanwar, S.~Racani\`ere et~al.,
  \emph{{Applications of flow models to the generation of correlated lattice
  QCD ensembles}},
  \href{https://doi.org/10.1103/PhysRevD.109.094514}{\emph{Phys. Rev. D}
  {\bfseries 109} (2024) 094514}
  [\href{https://arxiv.org/abs/2401.10874}{{\ttfamily 2401.10874}}].

\bibitem{Abbott:2022zsh}
R.~Abbott et~al., \emph{{Aspects of scaling and scalability for flow-based
  sampling of lattice QCD}},
  \href{https://doi.org/10.1140/epja/s10050-023-01154-w}{\emph{Eur. Phys. J. A}
  {\bfseries 59} (2023) 257}
  [\href{https://arxiv.org/abs/2211.07541}{{\ttfamily 2211.07541}}].

\bibitem{Komijani:2023fzy}
J.~Komijani and M.~K. Marinkovic, \emph{{Generative models for scalar field
  theories: how to deal with poor scaling?}},
  \href{https://doi.org/10.22323/1.430.0019}{\emph{PoS} {\bfseries LATTICE2022}
  (2023) 019} [\href{https://arxiv.org/abs/2301.01504}{{\ttfamily
  2301.01504}}].

\bibitem{Jarzynski:1996oqb}
C.~Jarzynski, \emph{{Nonequilibrium Equality for Free Energy Differences}},
  \href{https://doi.org/10.1103/PhysRevLett.78.2690}{\emph{Phys. Rev. Lett.}
  {\bfseries 78} (1997) 2690}
  [\href{https://arxiv.org/abs/cond-mat/9610209}{{\ttfamily
  cond-mat/9610209}}].

\bibitem{Jarzynski:1997ef}
C.~Jarzynski, \emph{{Equilibrium free-energy differences from nonequilibrium
  measurements: A master-equation approach}},
  \href{https://doi.org/10.1103/PhysRevE.56.5018}{\emph{Phys. Rev.} {\bfseries
  E56} (1997) 5018} [\href{https://arxiv.org/abs/cond-mat/9707325}{{\ttfamily
  cond-mat/9707325}}].

\bibitem{Jarzynski:1998ef}
C.~Jarzynski, \emph{{Equilibrium Free Energies from Nonequilibrium Processes}},
  {\emph{Acta Phys. Polon.} {\bfseries B29} (1998) 1609}
  [\href{https://arxiv.org/abs/cond-mat/9802155}{{\ttfamily
  cond-mat/9802155}}].

\bibitem{Crooks:1998}
G.~E. {Crooks}, \emph{{Nonequilibrium Measurements of Free Energy Differences
  for Microscopically Reversible Markovian Systems}},
  \href{https://doi.org/10.1023/A:1023208217925}{\emph{Journal of Statistical
  Physics} {\bfseries 90} (1998) 1481}.

\bibitem{Crooks_1999}
G.~E. Crooks, \emph{{Entropy production fluctuation theorem and the
  nonequilibrium work relation for free energy differences}},
  \href{https://doi.org/10.1103/physreve.60.2721}{\emph{Phys. Rev. E}
  {\bfseries 60} (1999) 2721}.

\bibitem{Caselle:2018kap}
M.~Caselle, A.~Nada and M.~Panero, \emph{{QCD thermodynamics from lattice
  calculations with nonequilibrium methods: The SU(3) equation of state}},
  \href{https://doi.org/10.1103/PhysRevD.98.054513}{\emph{Phys. Rev. D}
  {\bfseries 98} (2018) 054513}
  [\href{https://arxiv.org/abs/1801.03110}{{\ttfamily 1801.03110}}].

\bibitem{Francesconi:2020fgi}
O.~Francesconi, M.~Panero and D.~Preti, \emph{{Strong coupling from
  non-equilibrium Monte Carlo simulations}},
  \href{https://doi.org/10.1007/JHEP07(2020)233}{\emph{JHEP} {\bfseries 07}
  (2020) 233} [\href{https://arxiv.org/abs/2003.13734}{{\ttfamily
  2003.13734}}].

\bibitem{Bulgarelli:2023ofi}
A.~Bulgarelli and M.~Panero, \emph{{Entanglement entropy from non-equilibrium
  Monte Carlo simulations}},
  \href{https://doi.org/10.1007/JHEP06(2023)030}{\emph{JHEP} {\bfseries 06}
  (2023) 030} [\href{https://arxiv.org/abs/2304.03311}{{\ttfamily
  2304.03311}}].

\bibitem{Bulgarelli:2024onj}
A.~Bulgarelli and M.~Panero, \emph{{Duality transformations and the
  entanglement entropy of gauge theories}},
  \href{https://doi.org/10.1007/JHEP06(2024)041}{\emph{JHEP} {\bfseries 06}
  (2024) 041} [\href{https://arxiv.org/abs/2404.01987}{{\ttfamily
  2404.01987}}].

\bibitem{Bulgarelli:2025riv}
A.~Bulgarelli, M.~Caselle, A.~Nada and M.~Panero, \emph{{Casimir effect in
  critical O(N) models from nonequilibrium Monte Carlo simulations}},
  \href{https://doi.org/10.1103/lcvl-dgv4}{\emph{Phys. Rev. E} {\bfseries 112}
  (2025) 064126} [\href{https://arxiv.org/abs/2505.20403}{{\ttfamily
  2505.20403}}].

\bibitem{Vadacchino:2024lob}
D.~Vadacchino, A.~Nada and C.~Bonanno, \emph{{Topological susceptibility of
  SU(3) pure-gauge theory from out-of-equilibrium simulations}},
  \href{https://doi.org/10.22323/1.466.0415}{\emph{PoS} {\bfseries LATTICE2024}
  (2025) 415} [\href{https://arxiv.org/abs/2411.00620}{{\ttfamily
  2411.00620}}].

\bibitem{wu2020stochastic}
H.~Wu, J.~K\"{o}hler and F.~Noe, \emph{{Stochastic Normalizing Flows}},  in
  \emph{Advances in Neural Information Processing Systems}, vol.~33,
  pp.~\href{https://dl.acm.org/doi/10.5555/3495724.3496222}{5933--5944}, 2020,
  \href{https://arxiv.org/abs/2002.06707}{{\ttfamily 2002.06707}}.

\bibitem{Caselle:2024ent}
M.~Caselle, E.~Cellini and A.~Nada, \emph{{Numerical determination of the width
  and shape of the effective string using Stochastic Normalizing Flows}},
  \href{https://doi.org/10.1007/JHEP02(2025)090}{\emph{JHEP} {\bfseries 02}
  (2025) 090} [\href{https://arxiv.org/abs/2409.15937}{{\ttfamily
  2409.15937}}].

\bibitem{Bulgarelli:2024yrz}
A.~Bulgarelli, E.~Cellini, K.~Jansen, S.~K\"uhn, A.~Nada, S.~Nakajima et~al.,
  \emph{{Flow-Based Sampling for Entanglement Entropy and the Machine Learning
  of Defects}},
  \href{https://doi.org/10.1103/PhysRevLett.134.151601}{\emph{Phys. Rev. Lett.}
  {\bfseries 134} (2025) 151601}
  [\href{https://arxiv.org/abs/2410.14466}{{\ttfamily 2410.14466}}].

\bibitem{Neal2001}
R.~M. Neal, \emph{Annealed importance sampling},
  \href{https://doi.org/10.1023/A:1008923215028}{\emph{Statistics and
  Computing} {\bfseries 11} (2001) 125}
  [\href{https://arxiv.org/abs/physics/9803008}{{\ttfamily physics/9803008}}].

\bibitem{Dai2022}
C.~Dai, J.~Heng, P.~Jacob and N.~Whiteley, \emph{An invitation to sequential
  monte carlo samplers},
  \href{https://doi.org/10.1080/01621459.2022.2087659}{\emph{Journal of the
  American Statistical Association} {\bfseries 117} (2022) 1587}
  [\href{https://arxiv.org/abs/2007.11936}{{\ttfamily 2007.11936}}].

\bibitem{arbel2021annealed}
M.~Arbel, A.~Matthews and A.~Doucet, \emph{Annealed flow transport monte
  carlo},  in \emph{Proceedings of the 38th International Conference on Machine
  Learning}, vol.~139 of \emph{Proceedings of Machine Learning Research},
  pp.~318--330, PMLR, 18--24 Jul, 2021,
  \href{https://arxiv.org/abs/2102.07501}{{\ttfamily 2102.07501}},
  \href{https://proceedings.mlr.press/v139/arbel21a.html}{https://proceedings.mlr.press/v139/arbel21a.html}.

\bibitem{Matthews:2022sds}
A.~Matthews, M.~Arbel, D.~J. Rezende and A.~Doucet, \emph{Continual repeated
  annealed flow transport {M}onte {C}arlo},  in \emph{Proceedings of the 39th
  International Conference on Machine Learning}, vol.~162 of \emph{Proceedings
  of Machine Learning Research}, pp.~15196--15219, PMLR, 17--23 Jul, 2022,
  \href{https://arxiv.org/abs/2201.13117}{{\ttfamily 2201.13117}},
  \href{https://proceedings.mlr.press/v162/matthews22a.html}{https://proceedings.mlr.press/v162/matthews22a.html}.

\bibitem{Albergo:2024trn}
M.~S. Albergo and E.~Vanden-Eijnden, \emph{{NETS}: A non-equilibrium transport
  sampler},  in \emph{Proceedings of the 42nd International Conference on
  Machine Learning}, vol.~267 of \emph{Proceedings of Machine Learning
  Research}, pp.~1026--1055, PMLR, 13--19 Jul, 2025,
  \href{https://arxiv.org/abs/2410.02711}{{\ttfamily 2410.02711}},
  \href{https://proceedings.mlr.press/v267/albergo25a.html}{https://proceedings.mlr.press/v267/albergo25a.html}.

\bibitem{Wang:2023exq}
L.~Wang, G.~Aarts and K.~Zhou, \emph{{Diffusion models as stochastic
  quantization in lattice field theory}},
  \href{https://doi.org/10.1007/JHEP05(2024)060}{\emph{JHEP} {\bfseries 05}
  (2024) 060} [\href{https://arxiv.org/abs/2309.17082}{{\ttfamily
  2309.17082}}].

\bibitem{Zhu:2024kiu}
Q.~Zhu, G.~Aarts, W.~Wang, K.~Zhou and L.~Wang, \emph{{Diffusion models for
  lattice gauge field simulations}},  in \emph{{38th conference on Neural
  Information Processing Systems}}, 10, 2024,
  \href{https://arxiv.org/abs/2410.19602}{{\ttfamily 2410.19602}}.

\bibitem{Aarts:2024rsl}
G.~Aarts, D.~E. Habibi, L.~Wang and K.~Zhou, \emph{{On learning higher-order
  cumulants in diffusion models}},
  \href{https://doi.org/10.1088/2632-2153/adc53a}{\emph{Mach. Learn. Sci.
  Tech.} {\bfseries 6} (2025) 025004}
  [\href{https://arxiv.org/abs/2410.21212}{{\ttfamily 2410.21212}}].

\bibitem{Zhu:2025pmw}
Q.~Zhu, G.~Aarts, W.~Wang, K.~Zhou and L.~Wang, \emph{{Physics-conditioned
  diffusion models for lattice gauge theory}},
  \href{https://doi.org/10.1007/JHEP03(2026)111}{\emph{JHEP} {\bfseries 03}
  (2026) 111} [\href{https://arxiv.org/abs/2502.05504}{{\ttfamily
  2502.05504}}].

\bibitem{Aarts:2025lpi}
G.~Aarts, D.~E. Habibi, L.~Wang and K.~Zhou, \emph{{Combining complex Langevin
  dynamics with score-based and energy-based diffusion models}},
  \href{https://doi.org/10.1007/JHEP12(2025)160}{\emph{JHEP} {\bfseries 12}
  (2025) 160} [\href{https://arxiv.org/abs/2510.01328}{{\ttfamily
  2510.01328}}].

\bibitem{Creutz:1987xi}
M.~Creutz, \emph{{Overrelaxation and Monte Carlo Simulation}},
  \href{https://doi.org/10.1103/PhysRevD.36.515}{\emph{Phys. Rev. D} {\bfseries
  36} (1987) 515}.

\bibitem{Creutz:1980zw}
M.~Creutz, \emph{{Monte Carlo Study of Quantized $SU(2)$ Gauge Theory}},
  \href{https://doi.org/10.1103/PhysRevD.21.2308}{\emph{Phys. Rev. D}
  {\bfseries 21} (1980) 2308}.

\bibitem{Kennedy:1985nu}
A.~D. Kennedy and B.~J. Pendleton, \emph{{Improved Heat Bath Method for Monte
  Carlo Calculations in Lattice Gauge Theories}},
  \href{https://doi.org/10.1016/0370-2693(85)91632-6}{\emph{Phys. Lett. B}
  {\bfseries 156} (1985) 393}.

\bibitem{Cabibbo:1982zn}
N.~Cabibbo and E.~Marinari, \emph{{A New Method for Updating $SU(N)$ Matrices
  in Computer Simulations of Gauge Theories}},
  \href{https://doi.org/10.1016/0370-2693(82)90696-7}{\emph{Phys. Lett. B}
  {\bfseries 119} (1982) 387}.

\bibitem{Campostrini:1988cy}
M.~Campostrini, A.~Di~Giacomo and H.~Panagopoulos, \emph{{The Topological
  Susceptibility on the Lattice}},
  \href{https://doi.org/10.1016/0370-2693(88)90526-6}{\emph{Phys. Lett. B}
  {\bfseries 212} (1988) 206}.

\bibitem{Vicari:2008jw}
E.~Vicari and H.~Panagopoulos, \emph{{$\theta$ dependence of $SU(N)$ gauge
  theories in the presence of a topological term}},
  \href{https://doi.org/10.1016/j.physrep.2008.10.001}{\emph{Phys. Rept.}
  {\bfseries 470} (2009) 93} [\href{https://arxiv.org/abs/0803.1593}{{\ttfamily
  0803.1593}}].

\bibitem{DiVecchia:1981aev}
P.~Di~Vecchia, K.~Fabricius, G.~C. Rossi and G.~Veneziano, \emph{{Preliminary
  Evidence for $U(1)_A$ Breaking in QCD from Lattice Calculations}},
  \href{https://doi.org/10.1016/0550-3213(81)90432-6}{\emph{Nucl. Phys. B}
  {\bfseries 192} (1981) 392}.

\bibitem{DiVecchia:1981hh}
P.~Di~Vecchia, K.~Fabricius, G.~Rossi and G.~Veneziano, \emph{{Numerical Checks
  of the Lattice Definition Independence of Topological Charge Fluctuations}},
  \href{https://doi.org/10.1016/0370-2693(82)91203-5}{\emph{Phys. Lett. B}
  {\bfseries 108} (1982) 323}.

\bibitem{Campostrini:1989dh}
M.~Campostrini, A.~Di~Giacomo, H.~Panagopoulos and E.~Vicari,
  \emph{{Topological Charge, Renormalization and Cooling on the Lattice}},
  \href{https://doi.org/10.1016/0550-3213(90)90077-Q}{\emph{Nucl. Phys. B}
  {\bfseries 329} (1990) 683}.

\bibitem{DElia:2003zne}
M.~D'Elia, \emph{{Field theoretical approach to the study of theta dependence
  in Yang-Mills theories on the lattice}},
  \href{https://doi.org/10.1016/S0550-3213(03)00311-0}{\emph{Nucl. Phys. B}
  {\bfseries 661} (2003) 139}
  [\href{https://arxiv.org/abs/hep-lat/0302007}{{\ttfamily hep-lat/0302007}}].

\bibitem{Alles:1996nm}
B.~Alles, M.~D'Elia and A.~Di~Giacomo, \emph{{Topological susceptibility at
  zero and finite T in SU(3) Yang-Mills theory}},
  \href{https://doi.org/10.1016/S0550-3213(97)00205-8}{\emph{Nucl. Phys. B}
  {\bfseries 494} (1997) 281}
  [\href{https://arxiv.org/abs/hep-lat/9605013}{{\ttfamily hep-lat/9605013}}].

\bibitem{Alles:1997qe}
B.~Alles, M.~D'Elia and A.~Di~Giacomo, \emph{{Topology at zero and finite $T$
  in $SU(2)$ Yang--Mills theory}},
  \href{https://doi.org/10.1016/S0370-2693(97)01059-9}{\emph{Phys. Lett. B}
  {\bfseries 412} (1997) 119}
  [\href{https://arxiv.org/abs/hep-lat/9706016}{{\ttfamily hep-lat/9706016}}].

\bibitem{Lucini:2004yh}
B.~Lucini, M.~Teper and U.~Wenger, \emph{{Topology of $SU(N)$ gauge theories at
  $T \simeq 0$ and $T \simeq T_c$}},
  \href{https://doi.org/10.1016/j.nuclphysb.2005.02.037}{\emph{Nucl. Phys. B}
  {\bfseries 715} (2005) 461}
  [\href{https://arxiv.org/abs/hep-lat/0401028}{{\ttfamily hep-lat/0401028}}].

\bibitem{Giusti:2007tu}
L.~Giusti, S.~Petrarca and B.~Taglienti, \emph{{$\theta$ dependence of the
  vacuum energy in the $SU(3)$ gauge theory from the lattice}},
  \href{https://doi.org/10.1103/PhysRevD.76.094510}{\emph{Phys. Rev. D}
  {\bfseries 76} (2007) 094510}
  [\href{https://arxiv.org/abs/0705.2352}{{\ttfamily 0705.2352}}].

\bibitem{Panagopoulos:2011rb}
H.~Panagopoulos and E.~Vicari, \emph{{The $4D$ $SU(3)$ gauge theory with an
  imaginary $\theta$ term}},
  \href{https://doi.org/10.1007/JHEP11(2011)119}{\emph{JHEP} {\bfseries 11}
  (2011) 119} [\href{https://arxiv.org/abs/1109.6815}{{\ttfamily 1109.6815}}].

\bibitem{Bonati:2013tt}
C.~Bonati, M.~D'Elia, H.~Panagopoulos and E.~Vicari, \emph{{Change of
  \ensuremath{\theta} Dependence in $4D$ $SU(N)$ Gauge Theories Across the
  Deconfinement Transition}},
  \href{https://doi.org/10.1103/PhysRevLett.110.252003}{\emph{Phys. Rev. Lett.}
  {\bfseries 110} (2013) 252003}
  [\href{https://arxiv.org/abs/1301.7640}{{\ttfamily 1301.7640}}].

\bibitem{Berkowitz:2015aua}
E.~Berkowitz, M.~I. Buchoff and E.~Rinaldi, \emph{{Lattice QCD input for axion
  cosmology}}, \href{https://doi.org/10.1103/PhysRevD.92.034507}{\emph{Phys.
  Rev. D} {\bfseries 92} (2015) 034507}
  [\href{https://arxiv.org/abs/1505.07455}{{\ttfamily 1505.07455}}].

\bibitem{Borsanyi:2015cka}
S.~Borsanyi, M.~Dierigl, Z.~Fodor, S.~Katz, S.~Mages, D.~Nogradi et~al.,
  \emph{{Axion cosmology, lattice QCD and the dilute instanton gas}},
  \href{https://doi.org/10.1016/j.physletb.2015.11.020}{\emph{Phys. Lett. B}
  {\bfseries 752} (2016) 175}
  [\href{https://arxiv.org/abs/1508.06917}{{\ttfamily 1508.06917}}].

\bibitem{Bonati:2018rfg}
C.~Bonati, M.~Cardinali and M.~D'Elia, \emph{{$\theta$ dependence in trace
  deformed $SU(3)$ Yang-Mills theory: a lattice study}},
  \href{https://doi.org/10.1103/PhysRevD.98.054508}{\emph{Phys. Rev. D}
  {\bfseries 98} (2018) 054508}
  [\href{https://arxiv.org/abs/1807.06558}{{\ttfamily 1807.06558}}].

\bibitem{Bonati:2019kmf}
C.~Bonati, M.~Cardinali, M.~D'Elia and F.~Mazziotti, \emph{{$\theta$-dependence
  and center symmetry in Yang-Mills theories}},
  \href{https://doi.org/10.1103/PhysRevD.101.034508}{\emph{Phys. Rev. D}
  {\bfseries 101} (2020) 034508}
  [\href{https://arxiv.org/abs/1912.02662}{{\ttfamily 1912.02662}}].

\bibitem{Bonati:2015vqz}
C.~Bonati, M.~D'Elia, M.~Mariti, G.~Martinelli, M.~Mesiti, F.~Negro et~al.,
  \emph{{Axion phenomenology and $\theta$-dependence from $N_f = 2+1$ lattice
  QCD}}, \href{https://doi.org/10.1007/JHEP03(2016)155}{\emph{JHEP} {\bfseries
  03} (2016) 155} [\href{https://arxiv.org/abs/1512.06746}{{\ttfamily
  1512.06746}}].

\bibitem{Petreczky:2016vrs}
P.~Petreczky, H.-P. Schadler and S.~Sharma, \emph{{The topological
  susceptibility in finite temperature QCD and axion cosmology}},
  \href{https://doi.org/10.1016/j.physletb.2016.09.063}{\emph{Phys. Lett. B}
  {\bfseries 762} (2016) 498}
  [\href{https://arxiv.org/abs/1606.03145}{{\ttfamily 1606.03145}}].

\bibitem{Frison:2016vuc}
J.~Frison, R.~Kitano, H.~Matsufuru, S.~Mori and N.~Yamada, \emph{{Topological
  susceptibility at high temperature on the lattice}},
  \href{https://doi.org/10.1007/JHEP09(2016)021}{\emph{JHEP} {\bfseries 09}
  (2016) 021} [\href{https://arxiv.org/abs/1606.07175}{{\ttfamily
  1606.07175}}].

\bibitem{Borsanyi:2016ksw}
S.~Borsanyi et~al., \emph{{Calculation of the axion mass based on
  high-temperature lattice quantum chromodynamics}},
  \href{https://doi.org/10.1038/nature20115}{\emph{Nature} {\bfseries 539}
  (2016) 69} [\href{https://arxiv.org/abs/1606.07494}{{\ttfamily 1606.07494}}].

\bibitem{Bonati:2018blm}
C.~Bonati, M.~D'Elia, G.~Martinelli, F.~Negro, F.~Sanfilippo and A.~Todaro,
  \emph{{Topology in full QCD at high temperature: a multicanonical approach}},
  \href{https://doi.org/10.1007/JHEP11(2018)170}{\emph{JHEP} {\bfseries 11}
  (2018) 170} [\href{https://arxiv.org/abs/1807.07954}{{\ttfamily
  1807.07954}}].

\bibitem{Burger:2018fvb}
F.~Burger, E.-M. Ilgenfritz, M.~P. Lombardo and A.~Trunin, \emph{{Chiral
  observables and topology in hot QCD with two families of quarks}},
  \href{https://doi.org/10.1103/PhysRevD.98.094501}{\emph{Phys. Rev. D}
  {\bfseries 98} (2018) 094501}
  [\href{https://arxiv.org/abs/1805.06001}{{\ttfamily 1805.06001}}].

\bibitem{Chen:2022fid}
{\scshape TWQCD} collaboration, Y.-C. Chen, T.-W. Chiu and T.-H. Hsieh,
  \emph{{Topological susceptibility in finite temperature QCD with physical
  (u/d,s,c) domain-wall quarks}},
  \href{https://doi.org/10.1103/PhysRevD.106.074501}{\emph{Phys. Rev. D}
  {\bfseries 106} (2022) 074501}
  [\href{https://arxiv.org/abs/2204.01556}{{\ttfamily 2204.01556}}].

\bibitem{Athenodorou:2022aay}
A.~Athenodorou, C.~Bonanno, C.~Bonati, G.~Clemente, F.~D'Angelo, M.~D'Elia
  et~al., \emph{{Topological susceptibility of N$_{f}$ = 2 + 1 QCD from
  staggered fermions spectral projectors at high temperatures}},
  \href{https://doi.org/10.1007/JHEP10(2022)197}{\emph{JHEP} {\bfseries 10}
  (2022) 197} [\href{https://arxiv.org/abs/2208.08921}{{\ttfamily
  2208.08921}}].

\bibitem{Butti:2025rnu}
P.~Butti, M.~Della~Morte, B.~J{\"a}ger, S.~Martins and J.~T. Tsang,
  \emph{{Comparison of smoothening flows for the topological charge in QCD-like
  theories}}, \href{https://doi.org/10.1103/53vh-wm6v}{\emph{Phys. Rev. D}
  {\bfseries 112} (2025) 014504}
  [\href{https://arxiv.org/abs/2504.10197}{{\ttfamily 2504.10197}}].

\bibitem{Berg:1981nw}
B.~Berg, \emph{{Dislocations and Topological Background in the Lattice $O(3)$
  $\sigma$ Model}},
  \href{https://doi.org/10.1016/0370-2693(81)90518-9}{\emph{Phys. Lett. B}
  {\bfseries 104} (1981) 475}.

\bibitem{Iwasaki:1983bv}
Y.~Iwasaki and T.~Yoshie, \emph{{Instantons and Topological Charge in Lattice
  Gauge Theory}},
  \href{https://doi.org/10.1016/0370-2693(83)91111-5}{\emph{Phys. Lett. B}
  {\bfseries 131} (1983) 159}.

\bibitem{Itoh:1984pr}
S.~Itoh, Y.~Iwasaki and T.~Yoshie, \emph{{Stability of Instantons on the
  Lattice and the Renormalized Trajectory}},
  \href{https://doi.org/10.1016/0370-2693(84)90609-9}{\emph{Phys. Lett. B}
  {\bfseries 147} (1984) 141}.

\bibitem{Teper:1985rb}
M.~Teper, \emph{{Instantons in the Quantized $SU(2)$ Vacuum: A Lattice Monte
  Carlo Investigation}},
  \href{https://doi.org/10.1016/0370-2693(85)90939-6}{\emph{Phys. Lett. B}
  {\bfseries 162} (1985) 357}.

\bibitem{Ilgenfritz:1985dz}
E.-M. Ilgenfritz, M.~Laursen, G.~Schierholz, M.~M{\"u}ller-Preussker and
  H.~Schiller, \emph{{First Evidence for the Existence of Instantons in the
  Quantized $SU(2)$ Lattice Vacuum}},
  \href{https://doi.org/10.1016/0550-3213(86)90265-8}{\emph{Nucl. Phys. B}
  {\bfseries 268} (1986) 693}.

\bibitem{Alles:2000sc}
B.~Alles, L.~Cosmai, M.~D'Elia and A.~Papa, \emph{{Topology in $2D$ $CP^{N-1}$
  models on the lattice: A Critical comparison of different cooling
  techniques}}, \href{https://doi.org/10.1103/PhysRevD.62.094507}{\emph{Phys.
  Rev. D} {\bfseries 62} (2000) 094507}
  [\href{https://arxiv.org/abs/hep-lat/0001027}{{\ttfamily hep-lat/0001027}}].

\bibitem{APE:1987ehd}
{\scshape APE} collaboration, M.~Albanese et~al., \emph{{Glueball Masses and
  String Tension in Lattice QCD}},
  \href{https://doi.org/10.1016/0370-2693(87)91160-9}{\emph{Phys. Lett. B}
  {\bfseries 192} (1987) 163}.

\bibitem{Morningstar:2003gk}
C.~Morningstar and M.~J. Peardon, \emph{{Analytic smearing of SU(3) link
  variables in lattice QCD}},
  \href{https://doi.org/10.1103/PhysRevD.69.054501}{\emph{Phys. Rev. D}
  {\bfseries 69} (2004) 054501}
  [\href{https://arxiv.org/abs/hep-lat/0311018}{{\ttfamily hep-lat/0311018}}].

\bibitem{Narayanan:2006rf}
R.~Narayanan and H.~Neuberger, \emph{{Infinite $N$ phase transitions in
  continuum Wilson loop operators}},
  \href{https://doi.org/10.1088/1126-6708/2006/03/064}{\emph{JHEP} {\bfseries
  03} (2006) 064} [\href{https://arxiv.org/abs/hep-th/0601210}{{\ttfamily
  hep-th/0601210}}].

\bibitem{Luscher:2010iy}
M.~L{\"u}scher, \emph{{Properties and uses of the Wilson flow in lattice QCD}},
  \href{https://doi.org/10.1007/JHEP08(2010)071,
  10.1007/JHEP03(2014)092}{\emph{JHEP} {\bfseries 08} (2010) 071}
  [\href{https://arxiv.org/abs/1006.4518}{{\ttfamily 1006.4518}}].

\bibitem{Luscher:2011bx}
M.~L{\"u}scher and P.~Weisz, \emph{{Perturbative analysis of the gradient flow
  in non-abelian gauge theories}},
  \href{https://doi.org/10.1007/JHEP02(2011)051}{\emph{JHEP} {\bfseries 02}
  (2011) 051} [\href{https://arxiv.org/abs/1101.0963}{{\ttfamily 1101.0963}}].

\bibitem{Lohmayer:2011si}
R.~Lohmayer and H.~Neuberger, \emph{{Continuous smearing of Wilson Loops}},
  \href{https://doi.org/10.22323/1.139.0249}{\emph{PoS} {\bfseries LATTICE2011}
  (2011) 249} [\href{https://arxiv.org/abs/1110.3522}{{\ttfamily 1110.3522}}].

\bibitem{Bonati:2014tqa}
C.~Bonati and M.~D'Elia, \emph{{Comparison of the gradient flow with cooling in
  $SU(3)$ pure gauge theory}},
  \href{https://doi.org/10.1103/PhysRevD.89.105005}{\emph{Phys. Rev. D}
  {\bfseries D89} (2014) 105005}
  [\href{https://arxiv.org/abs/1401.2441}{{\ttfamily 1401.2441}}].

\bibitem{Alexandrou:2015yba}
C.~Alexandrou, A.~Athenodorou and K.~Jansen, \emph{{Topological charge using
  cooling and the gradient flow}},
  \href{https://doi.org/10.1103/PhysRevD.92.125014}{\emph{Phys. Rev. D}
  {\bfseries 92} (2015) 125014}
  [\href{https://arxiv.org/abs/1509.04259}{{\ttfamily 1509.04259}}].

\bibitem{JARTOP}
{C. Bonanno, A. Bulgarelli, E. Cellini, A. Nada, D. Panfalone, D. Vadacchino,
  L. Verzichelli}, \emph{\texttt{yang\_mills\_SNF\_Jarzynski}}. CPU code,
  \href{https://github.com/alessandronada/yang_mills_SNF_Jarzynski}{https://github.com/alessandronada/yang\_mills\_SNF\_Jarzynski}.

\bibitem{PTBC}
{C. Bonanno}, \emph{\texttt{yang\_mills\_PTBC}}. CPU code,
  \href{https://github.com/Claudio-Bonanno-93/yang_mills_PTBC}{https://github.com/Claudio-Bonanno-93/yang\_mills\_PTBC}.

\bibitem{yangmills}
{C. Bonati}, \emph{\texttt{yang\_mills}}. CPU code,
  \href{https://github.com/claudio-bonati/yang-mills}{https://claudio-bonati/yang-mills}.

\bibitem{Sivak_2012}
D.~A. Sivak and G.~E. Crooks, \emph{Thermodynamic metrics and optimal paths},
  \href{https://doi.org/10.1103/physrevlett.108.190602}{\emph{Phys. Rev. Lett.}
  {\bfseries 108} (2012) } [\href{https://arxiv.org/abs/1201.4166}{{\ttfamily
  1201.4166}}].

\bibitem{Jarzynski_2006}
C.~Jarzynski, \emph{Rare events and the convergence of exponentially averaged
  work values}, \href{https://doi.org/10.1103/physreve.73.046105}{\emph{Phys.
  Rev. E} {\bfseries 73} (2006) }
  [\href{https://arxiv.org/abs/cond-mat/0603185}{{\ttfamily
  cond-mat/0603185}}].

\bibitem{Vaikuntanathan_2011}
S.~Vaikuntanathan and C.~Jarzynski, \emph{{Escorted free energy simulations}},
  \href{https://doi.org/10.1063/1.3544679}{\emph{The Journal of Chemical
  Physics} {\bfseries 134} (2011) }
  [\href{https://arxiv.org/abs/1101.2612}{{\ttfamily 1101.2612}}].

\bibitem{Taco:2016equiv}
T.~Cohen and M.~Welling, \emph{Group equivariant convolutional networks},  in
  \emph{Proceedings of The 33rd International Conference on Machine Learning},
  vol.~48 of \emph{Proceedings of Machine Learning Research}, pp.~2990--2999,
  PMLR, 20--22 Jun, 2016, \href{https://arxiv.org/abs/1602.07576}{{\ttfamily
  1602.07576}},
  \href{https://proceedings.mlr.press/v48/cohenc16.html}{https://proceedings.mlr.press/v48/cohenc16.html}.

\bibitem{kohler2019equivariant}
J.~K{\"o}hler, L.~Klein and F.~No{\'e}, \emph{Equivariant flows: sampling
  configurations for multi-body systems with symmetric energies},
  \href{https://arxiv.org/abs/1910.00753}{{\ttfamily 1910.00753}}.

\bibitem{Nagai:2021bhh}
Y.~Nagai and A.~Tomiya, \emph{{Gauge covariant neural network for quarks and
  gluons}}, \href{https://doi.org/10.1103/PhysRevD.111.074501}{\emph{Phys. Rev.
  D} {\bfseries 111} (2025) 074501}
  [\href{https://arxiv.org/abs/2103.11965}{{\ttfamily 2103.11965}}].

\bibitem{SNFSU3}
{A. Bulgarelli, E. Cellini, A. Nada}, \emph{\texttt{snf\_su3}}. CPU/GPU PyTorch
  code,
  \href{https://github.com/alessandronada/snf_su3}{https://github.com/alessandronada/snf\_su3}.

\bibitem{Kingma:2014vow}
D.~P. Kingma and J.~Ba, \emph{{Adam: A Method for Stochastic Optimization}},
  \href{https://arxiv.org/abs/1412.6980}{{\ttfamily 1412.6980}}.

\bibitem{Necco:2001xg}
S.~Necco and R.~Sommer, \emph{{The $N_{\rm f}=0$ heavy quark potential from
  short to intermediate distances}},
  \href{https://doi.org/10.1016/S0550-3213(01)00582-X}{\emph{Nucl. Phys. B}
  {\bfseries 622} (2002) 328}
  [\href{https://arxiv.org/abs/hep-lat/0108008}{{\ttfamily hep-lat/0108008}}].

\bibitem{Bonanno:2025eeb}
C.~Bonanno, \emph{{The large-$N$ limit of the topological susceptibility of
  $\mathrm{SU}(N)$ Yang-Mills theories via Parallel Tempering on Boundary
  Conditions}}, \href{https://doi.org/10.1007/JHEP01(2026)039}{\emph{JHEP}
  {\bfseries 01} (2026) 039}
  [\href{https://arxiv.org/abs/2510.08006}{{\ttfamily 2510.08006}}].

\bibitem{Schmiedl_2007}
T.~Schmiedl and U.~Seifert, \emph{Optimal finite-time processes in stochastic
  thermodynamics},
  \href{https://doi.org/10.1103/physrevlett.98.108301}{\emph{Phys. Rev. Lett.}
  {\bfseries 98} (2007) }
  [\href{https://arxiv.org/abs/cond-mat/0701554}{{\ttfamily
  cond-mat/0701554}}].

\bibitem{Gomez_Marin_2008}
A.~Gomez-Marin, T.~Schmiedl and U.~Seifert, \emph{Optimal protocols for minimal
  work processes in underdamped stochastic thermodynamics},
  \href{https://doi.org/10.1063/1.2948948}{\emph{The Journal of Chemical
  Physics} {\bfseries 129} (2008) }
  [\href{https://arxiv.org/abs/0803.0269}{{\ttfamily 0803.0269}}].

\bibitem{Zulkowski_2012}
P.~R. Zulkowski, D.~A. Sivak, G.~E. Crooks and M.~R. DeWeese, \emph{Geometry of
  thermodynamic control},
  \href{https://doi.org/10.1103/physreve.86.041148}{\emph{Phys. Rev. E}
  {\bfseries 86} (2012) } [\href{https://arxiv.org/abs/1208.4553}{{\ttfamily
  1208.4553}}].

\bibitem{Rotskoff_2015}
G.~M. Rotskoff and G.~E. Crooks, \emph{Optimal control in nonequilibrium
  systems: Dynamic riemannian geometry of the ising model},
  \href{https://doi.org/10.1103/physreve.92.060102}{\emph{Phys. Rev. E}
  {\bfseries 92} (2015) } [\href{https://arxiv.org/abs/1510.06734}{{\ttfamily
  1510.06734}}].

\bibitem{Blaber_2020}
S.~Blaber and D.~A. Sivak, \emph{Skewed thermodynamic geometry and optimal free
  energy estimation}, \href{https://doi.org/10.1063/5.0033405}{\emph{The
  Journal of Chemical Physics} {\bfseries 153} (2020) }
  [\href{https://arxiv.org/abs/2009.14354}{{\ttfamily 2009.14354}}].

\bibitem{Blaber_2021}
S.~Blaber, M.~D. Louwerse and D.~A. Sivak, \emph{Steps minimize dissipation in
  rapidly driven stochastic systems},
  \href{https://doi.org/10.1103/physreve.104.l022101}{\emph{Phys. Rev. E}
  {\bfseries 104} (2021) } [\href{https://arxiv.org/abs/2105.04691}{{\ttfamily
  2105.04691}}].

\bibitem{Bonanca_2018}
M.~V.~S. Bonança and S.~Deffner, \emph{Minimal dissipation in processes far
  from equilibrium},
  \href{https://doi.org/10.1103/physreve.98.042103}{\emph{Phys. Rev. E}
  {\bfseries 98} (2018) } [\href{https://arxiv.org/abs/1803.07050}{{\ttfamily
  1803.07050}}].

\bibitem{Kamizaki_2022}
L.~P. Kamizaki, M.~V.~S. Bonança and S.~R. Muniz, \emph{Performance of optimal
  linear-response processes in driven brownian motion far from equilibrium},
  \href{https://doi.org/10.1103/physreve.106.064123}{\emph{Phys. Rev. E}
  {\bfseries 106} (2022) } [\href{https://arxiv.org/abs/2204.07145}{{\ttfamily
  2204.07145}}].

\bibitem{engel2023optimal}
M.~C. Engel, J.~A. Smith and M.~P. Brenner, \emph{Optimal control of
  nonequilibrium systems through automatic differentiation},
  \href{https://doi.org/10.1103/PhysRevX.13.041032}{\emph{Phys. Rev. X}
  {\bfseries 13} (2023) 041032}
  [\href{https://arxiv.org/abs/2201.00098}{{\ttfamily 2201.00098}}].

\bibitem{Blaber_2023}
S.~Blaber and D.~A. Sivak, \emph{Optimal control in stochastic thermodynamics},
  \href{https://doi.org/10.1088/2399-6528/acbf04}{\emph{J. Phys. Commun.}
  {\bfseries 7} (2023) 033001}
  [\href{https://arxiv.org/abs/2212.00706}{{\ttfamily 2212.00706}}].

\end{thebibliography}\endgroup

\end{document}